%% file: manuscript.tex
\documentclass[%
superscriptaddress,
twocolumn,
groupedaddress,
amsmath,amssymb,
aps,
prb,
floatfix,
showkeys,
reprint,
]{revtex4-1}

\usepackage{graphicx}
\usepackage{dcolumn}
\usepackage{bm}
\usepackage{amsmath}
\usepackage{amssymb}
\usepackage{multirow}
\usepackage{xcolor}
\usepackage{booktabs}
\usepackage{tabularx}
\usepackage[english]{babel}
\usepackage[utf8]{inputenc}
\usepackage{enumerate}
\usepackage{newunicodechar}
\newunicodechar{ﬁ}{fi}
\newunicodechar{ﬀ}{ff}

\newcommand{\etal}{\textit{et al.\/}}
\newcommand{\ie}{i.\,e.,}

\newcommand{\uB}{\ensuremath{\mu_{\text{B}}}}

\begin{document}
\title{Electronic and magnetic properties of 3$d$ transition-metal adatoms on Mn/W(110)}

\author{Mara Gutzeit}
\affiliation{Institute of Theoretical Physics and Astrophysics, University of Kiel, Leibnizstrasse 15, 24098 Kiel, Germany}

\author{Soumyajyoti Haldar}
\email[Corresponding author: ]{haldar@physik.uni-kiel.de\\haldar.physics@gmail.com}
\affiliation{Institute of Theoretical Physics and Astrophysics, University of Kiel, Leibnizstrasse 15, 24098 Kiel, Germany}

\author{Stefan Heinze}
\affiliation{Institute of Theoretical Physics and Astrophysics, University of Kiel, Leibnizstrasse 15, 24098 Kiel, Germany}

\date{\today}

\begin{abstract}
Using density functional theory, we investigate the electronic and magnetic properties of $3d$ transition-metal adatoms adsorbed on a monolayer of Mn on W(110). Mn/W(110) has a noncollinear cycloidal spin-spiral ground state with an angle of 173$^\circ$ between magnetic moments of adjacent Mn rows. It allows to rotate the spin orientation of an adsorbed magnetic adatom quasi-continuously. Therefore, this surface is ideally suited for manipulating the spin direction of individual atoms and exploring  their magnetic properties using scanning tunneling microscopy (STM). The adsorbed V and Cr transition-metal adatoms couple antiferromagnetically to the nearest neighbor Mn atom of Mn monolayer while Mn, Fe, Co, and Ni couple ferromagnetically. The magnetic moments of the $3d$  adatoms are large and show a Hund's rule type of trend with a peak in the middle of the series. We find large spin splitting of the $3d$ transition-metal adatoms, large spin polarization of the local vacuum density of states up to 73\% at the Fermi energy, and significant tunneling anisotropic magnetoresistance enhancement up to 27\%. We conclude that such  large values stem from the strong hybridization between the adatoms and the Mn atoms of the monolayer.  Furthermore, identification of spin orientations of the adatom using spin-polarized STM is only possible for Co and V adatoms. 
\end{abstract}

\maketitle

\section{Introduction}
In the field of spintronics, noncollinear magnetic structures, such as chiral domain walls~\cite{Heide2008,Ryu2013,Emori2013}, skyrmions~\cite{Bogdanov1989,Bogdanov:2001aa,Heinze2011,Romming2013,Herve2018}, and spin spirals~\cite{Ferriani2008,Phark2014} are being investigated extensively by many research groups due to their fascinating dynamical and transport properties~\cite{Fert2013,Nagaosa2013}. 
A monolayer Mn grown on W(110) [Mn/W(110)] is a very prominent example for a cycloidal spin-spiral ground state which is driven by the Dzyaloshinskii-Moriya interaction~\cite{Bode2007}. 
Another exciting research direction is
single magnetic adatoms and nanostructures on such noncollinear surfaces~\cite{Lounis2007,Enders2010,Hermenau2017}. For example, using the surface of Mn/W(110) with a noncollinear magnetic structure, the spin orientation of an adsorbed adatom can be controlled due to local exchange coupling without the presence of an external magnetic field.
Thereby, the transport properties of magnetic adatoms as a function of spin direction can be probed using spin-polarized scanning tunneling microscopy (SP-STM)~\cite{Wiesendanger1990,Wulfhekel1999,Bode2003}.
Recently, even a simultaneous measurement of the spin-polarized tunneling current and of the local exchange force between a magnetic tip and a surface with a noncollinear spin structure has been obtained~\cite{Hauptmann2017,Hauptmann2019}. 

Using SP-STM, Serrate {\etal} have imaged the spin direction of individual Co adatoms adsorbed on Mn/W(110)~\cite{Serrate2010,Serrate2016}. 
These experiments show that by means of SP-STM, the spin direction of the Co adatom can be controlled while maintaining the magnetic sensitivity by laterally moving the adatom along the Mn/W(110) surface. 
In a recent theoretical investigation~\cite{Haldar2018}, the authors explained the shape asymmetry observed in the SP-STM by Serrate {\etal}~\cite{Serrate2010,Serrate2016} and showed that the noncollinear spin states of the Mn layers of Mn/W(110) are also reflected in the orbitals of the adsorbed Co adatom due to strong hybridization between the $3d$ orbitals of the Co adatom and Mn atoms of the surface. 
It is an interesting question whether the characteristic features observed for Co adatoms adsorbed on Mn/W(110) can also be seen for other $3d$ transition-metal (TM) adatoms. However, such investigations, both theoretically and experimentally,  are missing.

One aspect of such studies will be to compare the structural, magnetic, and electronic properties of various $3d$ TM adatoms adsorbed on Mn/W(110). It will be interesting to understand the origin of spin-polarized electrons observed near the Fermi energy ($E_F$) in SP-STM experiments for other $3d$ TM adatoms. The origin of spin-polarized electrons are based on two main arguments~\cite{Ferriani2010}: (i) one would expect $d$ electrons of the $3d$ TM adatoms to dominate at the apex atom of the tip. (ii) Another argument is that in the vacuum region the local density of states (LDOS), the prominent quantity for the tunneling current, can be dominated by more de-localized $s$ or $p$ electrons. The spin orientation of Co adatoms on Mn/W(110) can be identified in SP-STM experiments due to the presence of majority and minority states of different orbital symmetry near $E_F$~\cite{Serrate2010,Haldar2018,Ferriani2010,Zhou2010}. Whether similar observations can be made for other $3d$ TM adatoms adsorption are subject of investigations.       

The other aspect of the experiment of Co adatoms on Mn/W(110) is the indirect evidence of tunneling anisotropic magnetoresistance (TAMR) which has been observed in the STM line profiles~\cite{Serrate2010,Serrate2016}.
The TAMR is very appealing for spintronic applications as it can be observed with only one ferromagnetic electrode and it does not need any coherent spin-dependent transport~\cite{Fert2008,Sinova2012}. 
TAMR has been observed in various systems, such as planar ferromagnetic surfaces~\cite{Shick2006,Chantis2007}, tunnel junctions~\cite{Gould2004,Matos2009a,Matos2009b,Gao2007}, mechanically controlled break junctions~\cite{Viret2006,Bolotin2006}. TAMR values of $\sim$ 10\% have been observed in these cases. 

Since TAMR is driven by spin-orbit coupling (SOC), researchers have been trying to increase the TAMR by using elements having significant SOC. N\'{e}el {\etal} have recently achieved $\approx$ 12\% TAMR by adsorbing Co adatom on a bi-layer Fe film grown on a W(110) substrate~\cite{Neel2013}. A TAMR of up to 30\% mediated by surface states has been reported recently for Co films on Ru(0001)~\cite{Herve2018b}. TAMR values of $\approx$ 20\% have also been achieved using suitably oriented Pb dimers on the Fe double layer on W(110)~\cite{Schoneberg2018}. Recently, we have predicted up to 60\% TAMR for Pb and Bi adatom and dimers adsorbed on Mn/W(110)~\cite{Haldar2019}. In an earlier theoretical study, Caffrey {\etal} have compared the electronic structure, magnetic properties, and TAMR for Co, Rh, and Ir adsorbed on Mn/W(110) and predicted $\approx$ 50\% TAMR values for the Ir adatom adsorption~\cite{Caffrey2014a}. However, there is no systematic study of TAMR for other $3d$ TM adatoms on Mn/W(110) available in the literature. 

Here, we explore the adsorption of various $3d$ TM adatoms on Mn/W(110) and compare their structural, magnetic, and electronic properties, spin polarization of the vacuum LDOS, and the TAMR effect using first-principles density functional theory (DFT) calculations. Our results for Co adatoms on Mn/W(110) match well with previously published results~\cite{Caffrey2014a,Serrate2010,Haldar2018}.  We have locally approximated the spin structure of Mn/W(110) as a two-dimensional antiferromagnet~\cite{Heinze2000}. Using the  local exchange interaction, the magnetization direction can be tuned quasi-continuously on this substrate with noncollinear magnetic ordering~\cite{Caffrey2014a}. However, for simplicity, we have considered two limiting cases of spin direction: (i) out-of-plane where the magnetization direction is perpendicular to the surface and (ii) in-plane  where the magnetization direction is pointing along the $[1\overline{1}0]$ direction. 
Our calculations show that V and Cr transition-metal adatoms couple antiferromagnetically to the nearest neighbor Mn atom of Mn monolayer while Mn, Fe, Co, and Ni couple ferromagnetically. The magnetic moments of the $3d$  adatoms are large and show a Hund's rule type of trend with a peak in the middle of the series. Large values of the spin polarization of the vacuum LDOS up to 73\% at $E_F$ and significant enhancement of TAMR values ranging from $\approx$ 17\% up to $\approx$ 27\% due to local enhancement of SOC are found due to strong hybridization. Furthermore, as for Co adatoms on Mn/W(110), our results show that the spin orientation of the respective adatom can only be distinguished in SP-STM experiments for V adatom.

This paper is organized as follows. First, we briefly discuss the computational methods for our calculations in section~\ref{sec:compdet}. Then we present the relaxed geometries and give a detailed analysis of the electronic structure and magnetic moments of the $3d$ TM adatoms on Mn/W(110) in section~\ref{subsec:str_mag}. Here, we particularly focus on the spin polarization in the vacuum LDOS which is amenable to experiments and hybridization effects with the substrate. Furthermore results for the TAMR of the adatoms will be presented as well as explained by means of a simplified model of two localized atomic surface states interacting via SOC (section~\ref{sec:tamr_model_result}). Finally, we summarize our main conclusions in section~\ref{sec:conclusion}.

\section{Computational details}
\label{sec:compdet}
\begin{figure}[htbp]
	\centering
	\includegraphics[scale=0.85,clip]{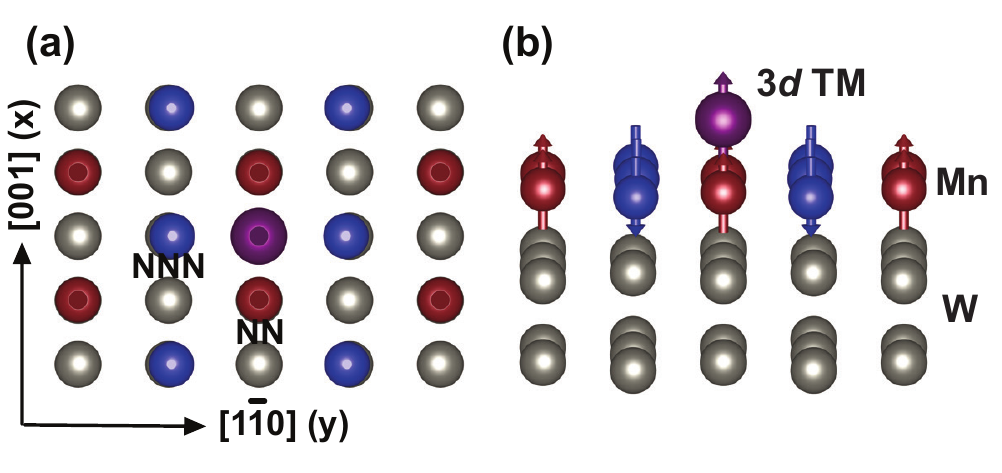}
	\caption{(a) Top view and (b) perspective view of the supercell used for the adatom on Mn/W(110) calculations. Gray spheres represent the W substrate while the Mn atoms are depicted as red (blue) spheres with arrows showing the approximated antiferromagnetic order with respect to the adatom (purple sphere). `NN' and `NNN' refer to the nearest neighbor and next nearest neighbor Mn atoms of the adatom, whereas $x$ and $y$ refer to the labeling of the $d$-orbitals.   }
	\label{fig:geom}
\end{figure}
In this work density functional theory (DFT) calculations have been performed within the projector augmented wave method (PAW)~\cite{blo,blo1} as implemented in the \textsc{vasp} code~\cite{vasp1,vasp2}. Exchange-correlation effects are taken into account within the  generalized gradient approximation (GGA) of Perdew-Burke-Ernzerhof
(PBE)~\cite{PBE,PBEerr}. An energy cutoff of 450 eV has been used to converge the plane wave basis set. Structural relaxations for the positions of the adatoms and the Mn atoms of the monolayer have been carried out using a $6\times6\times1$ k-point Monkhorst-Pack mesh~\cite{Monkhorst1976} and with a force tolerance of $10^{-2}$ eV /{\AA}. The effect of SOC has been included within the noncollinear mode of \textsc{vasp} as implemented by Hobbs {\etal}~\cite{Hobbs2000}.
Here, 400 k$_\parallel$-points in the full two-dimensional Brillouin zone are used for computing the LDOS directly at the adatom as well as in the vacuum. The density of states in the vacuum was calculated by placing an empty sphere at a specific height above the adatom onto which the density of states was projected.    

The DFT calculations for the Mn/W(110) system have been carried out using a symmetric slab consisting of five atomic layers of W with a pseudomorphic Mn layer - as experimentally observed~\cite{Bode1998} - on each side. 
We have used the GGA lattice constant of W (3.17 {\AA}) for our calculations as it agrees well with the experimental value of 3.165 {\AA}. 
We have approximately considered the local magnetic order of the system collinear due to the long periodicity of the spin spiral ground state of the Mn/W(110) substrate.
Therefore, we approximate the local magnetic order as a two-dimensional antiferromagnet (AFM) using a $c(4\times4)$ AFM surface unit cell, as shown in Figs.~\ref{fig:geom}(a) and~\ref{fig:geom}(b). The respective adatom is centered in the hollow-site position within the unit cell on both sides of the slab. The minimum distance between the periodic images of the adatom is 6.34 {\AA}, which is large enough to keep the interactions between the periodic adatoms negligible.
In the case of V and Cr adatoms, we have used 
a larger $c(6\times6)$ AFM surface unit cell as the $d$ orbitals of these elements have a larger radial extent due to the smaller attractive Coulomb potential of the nucleus. 
The minimum distance between the periodic images of the adatoms amounts to 9.51 {\AA} in these cases.
Calculations of the adsorption of a Mn atom have shown that this element represents a limiting case in which the smaller $c(4\times4)$ AFM surface unit cell is still sufficient to keep interactions between periodic adatoms negligible.
In addition, a thick, approximately 25 {\AA}, vacuum layer is included in the $z$ direction normal to the surface to remove interactions between periodic slabs. 

The binding energy $E_b$ of a $3d$ adatom adsorbed on Mn/W(110) is defined as
\begin{equation}
\label{eq:binding}
\begin{split}
E_b & = [\; \mathrm{N} * E(3d) + E(Mn/W(110)) \\ 
& - E(3d@Mn/W(110)) \;] \;/\; \mathrm{N}
\end{split}
\end{equation}
where $E(3d@Mn/W(110))$ and  $E(Mn/W(110))$ are the total energies of the optimized Mn/W(110) surface with and without $3d$ adatom adsorption, respectively. $E(3d)$ is the energy of an isolated $3d$ adatom placed in a cubic cell of 30 {\AA} (only the $\Gamma$ point of the Brillouin zone was sampled in this case). N is the number of adatoms in the supercell. Here, N = 2, as we have used a symmetric slab.

\section{Results and discussion}
\subsection{Structural and magnetic properties}
\label{subsec:str_mag}
\begin{figure*}[htbp]
\centering
\includegraphics[scale=0.9]{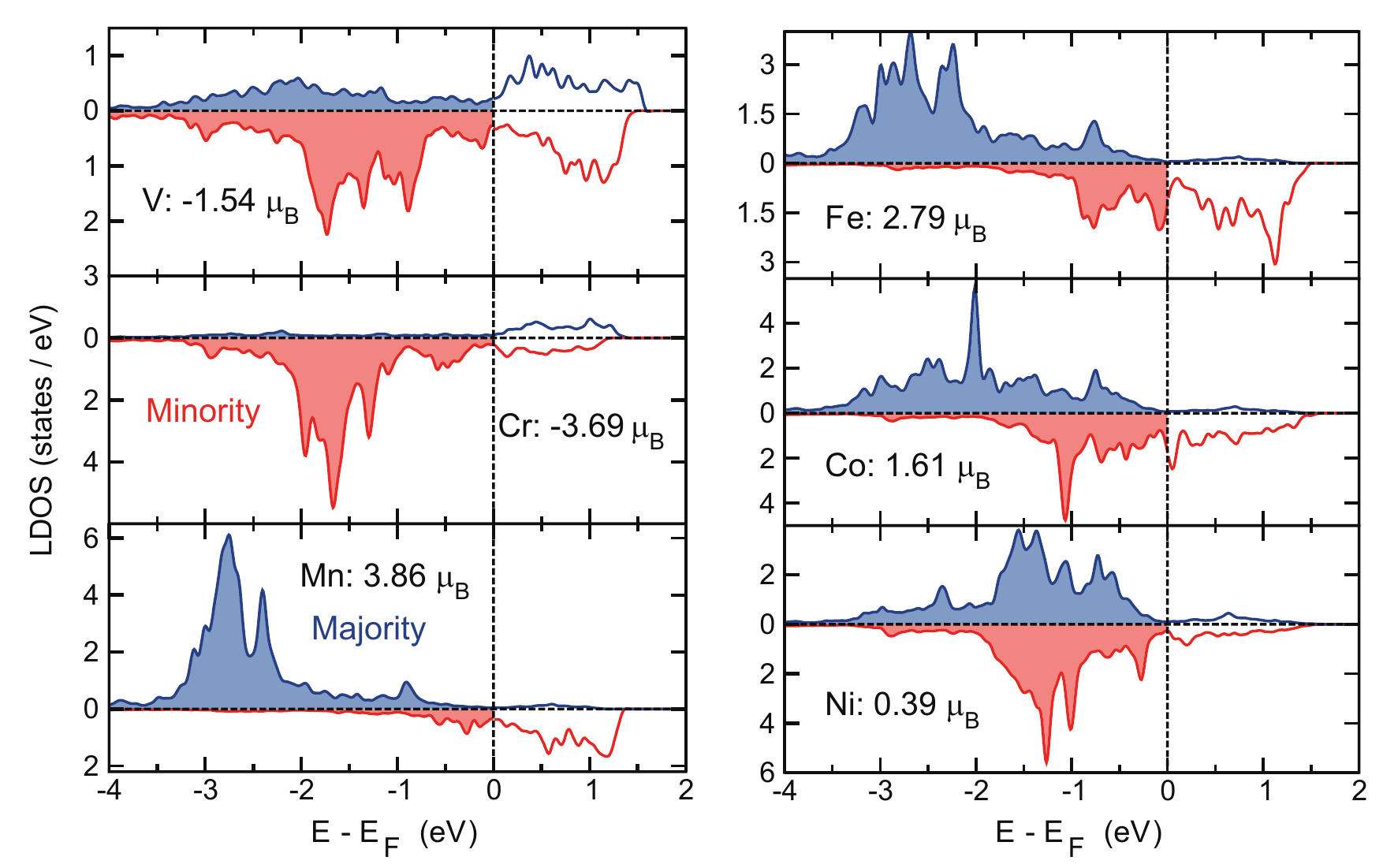}
\caption{Comparison of the scalar relativistic calculated local density of states (LDOS) of all studied $3d$ transition-metal adatoms on Mn/W(110). The majority states which are defined with respect to the nearest neighbor Mn atoms in the Mn monolayer are depicted in blue color, the minority states in red color, respectively. The magnetic moments of each $3d$ TM adatom are noted additionally in order to relate them to the exchange splitting of the spin channels.}
\label{fig:3ddos}
\end{figure*}
\begin{table}[htb]
	\centering
	\caption{Binding energy per adatom [$E_b$] (in eV), distance (in {\AA}) of $3d$ transition-metal adatoms from nearest neighbor (NN) [$d_{\text{NN}}$], next-nearest neighbor (NNN) [$d_{\text{NNN}}$] Mn atoms of the Mn monolayer on W(110) and from the Mn monolayer [$d_{\text{Z}}$]. $\Delta$x, $\Delta$y and $\Delta$z specify the displacement with respect to the clean surface of the NN and NNN Mn atoms after adsorption of the respective adatom. Positive (negative) values imply that the Mn atoms move towards (away from) the adatom. The calculated distances for the Co adatom match well with the previously published result of Caffrey {\etal}~\cite{Caffrey2014a}.~\\}
	\label{tab:table1}
	\begin{ruledtabular}
		\begin{tabular}{c c c c c c c }
			& V & Cr & Mn & Fe & Co & Ni\\ 
			\colrule
			$E_b$ & 4.25  & 2.69  & 2.80 & 3.71  & 4.10 & 4.31 \\
			$d_{\text{NN}}$	&2.09  & 2.36  & 2.50 &2.33  & 2.26 & 2.30 \\
			$d_{\text{NNN}}$	&2.77 & 2.89  & 2.69 & 2.62 & 2.61 & 2.63 \\ 
			$d_{\text{Z}}$ &1.56 & 1.80 & 1.84 & 1.68 & 1.60 & 1.64 \\
			$\Delta$z$_{\text{NN}}$	&$-0.03$  &$-0.03$  & $-0.14$ &$-0.12$  & $-0.08$ & $-0.06$ \\ 
			$\Delta$z$_{\text{NNN}}$	& +0.01 &+0.01 &+0.05 &+0.04 &+0.07 & +0.09 \\
			$\Delta$x$_{\text{NN}}$&+0.21 &+0.08 & $-0.10$&+0.01 & +0.02 & $-0.02$ \\ 
			$\Delta$y$_{\text{NNN}}$&$-0.06$ &$-0.04$&+0.12 &+0.11 & +0.09 & +0.09 \\
		\end{tabular} 
	\end{ruledtabular}
\end{table}
We start our discussion with the modification of structural and magnetic properties of the Mn/W(110) surface due to the adsorption of individual adatoms from the $3d$ TM series (see Table~\ref{tab:table1}). 

A noticeable buckling of the Mn monolayer in the vicinity of the adsorption site can be observed due to structural relaxations. 
After the adsorption of $3d$ TM adatoms, the nearest neighbor (NN) Mn atoms move away from the adsorbate. 
These movements are relatively larger for the adatoms from the middle of the $3d$ series (Mn and Fe) as compared to the elements either beginning or end of the $3d$ series (V and Ni). In contrast, the displacements of the next-nearest neighbor (NNN) Mn atoms, which move slightly towards the respective $3d$ TM adatoms, are small.
However, we found no general trend concerning the displacements of the atomic positions in the Mn monolayer along the $3d$ series. In addition it can be seen that
both the distances between the adatom and the NN Mn atoms as well as their distance from the surface become maximal in the middle of the series (for the Mn adatom) and decrease again for the other adatoms. This effect can be explained by means of the covalent radius which is largest for the Mn adatom and by means of the binding energy with the underlying surface. As it can be seen from Table~\ref{tab:table1}, the binding energy becomes minimal for the Cr and Mn adatom which 
is the consequence of a weak metallic bond with the Mn monolayer due to nearly half-filling of the $3d$ 
band and large exchange splitting.


\begin{table}[htb]
\centering
\caption{Spin magnetic moments (in  {\uB}) of the adsorbed 3$d$ transition-metal adatoms ($\mu_{\mathrm{spin}}$) as well as the nearest neighbor (NN) and next-nearest neighbor (NNN) Mn atoms of the Mn monolayer on W(110). For comparison, the spin magnetic moment of the Mn atoms from the clean Mn/W(110) surface is also tabulated. The calculated spin moment values for the Co adatom match well with the previously published result of Caffrey {\etal}~\cite{Caffrey2014a}. The orbital moments $\mu_{\mathrm{orb}}$ 
(in {\uB}) of the adsorbed $3d$ adatoms are also tabulated for completeness.~\\ }
\label{tab:table2}
\begin{ruledtabular}
\begin{tabular}{c c c c c c c} 
 	& V & Cr & Mn & Fe & Co & Ni\\ 
	\colrule 
$\mu_{\mathrm{spin}}$	& $-$1.54 & $-$3.69  & +3.86 & +2.79 & +1.61 & +0.39 \\ 
Mn$_\text{NN}$	& +1.82 & +2.84 & +2.33 & +2.28 & +2.59 & +2.93\\ 
Mn$_\text{NNN}$	& $-$3.19 & $-$3.28 &$-3.36$ &$-3.25$ &$-3.33$ & $-3.43$\\
Mn$_\text{clean}$ & $\pm$3.41& $\pm$3.41& $\pm$3.41& $\pm$3.41& $\pm$3.41 & $\pm$3.41 \\
$\mu_{\mathrm{orb}}$ & 0.03 & 0.02  & 0.01 & 0.10 & 0.12 & 0.04\\
\end{tabular} 
\end{ruledtabular}
\end{table}
The magnetic moments of the $3d$ transition-metal adatoms, the NN Mn atoms, and the NNN Mn atoms after the adsorption of the $3d$ TM adatoms on Mn/W(110) are given in Table~\ref{tab:table2}. 
The magnetic moments of Mn atoms of the clean Mn/W(110) surface are $\pm$3.41 {\uB}. 
The magnetic moments of the $3d$ adatoms display a Hund's rule like trend of an increase towards the middle of the series and a decrease towards the end as observed for such adatoms on other surfaces.  
The moment of the Mn monolayer is reduced after the adsorption of the adatoms due to the hybridization between NN Mn atoms and the respective adatom.  
In all cases a significant reduction of the magnetic moment can be observed for the NN Mn atoms, 
in particular for the V adsorbate where it is reduced by 1.6 {\uB}. The V adatom itself, due to the hybridization with underlying Mn layer, carries a magnetic moment of $-1.54$ {\uB} thereby coupling antiferromagnetically (AFM) with the NN Mn atoms and ferromagnetically (FM) with the NNN Mn atoms (see Table~\ref{tab:table2}). The same holds true for the Cr adatom, whereas for the rest of the $3d$ TM series a FM coupling between the respective adatom and the NN Mn atoms of the Mn monolayer is favored by the exchange interaction.
It is worth mentioning that an AFM alignment with the NN Mn atoms could also be stabilized in the case of Fe and Co adsorption. While our calculations reveal a very small energetic difference of 6 meV between FM and AFM state for the Fe adatom, this difference is significantly larger for the Co adatom with 108 meV.
This large value matches well with previous reports of the Co adatom on Mn/W(110)~\cite{Caffrey2014a, Serrate2010}. 
A possible FM coupling between the Cr adsorbate and the NN Mn atoms can also be stabilized. However, this state is energetically unfavorable by 197 meV per Cr adatom as compared to the case of an AFM alignment.

The reduction of magnetic moment can also be seen for the NNN Mn atoms, albeit the effect of hybridization is less pronounced in these cases. The largest drop for the magnetic moment occurs for V adatom adsorption where the magnetic moments of the NNN Mn atoms are reduced by 0.22 {\uB}. It matches with the trends observed in relaxations, where large hybridization due to larger extent of $3d$ orbitals occurs at the beginning (V, Cr) of $3d$ series and become smaller at the end (Ni).

Figure~\ref{fig:3ddos} shows a comparison of the LDOS of the $3d$ orbitals of the adatoms. The calculations show a good agreement of the exchange splitting of the majority and minority bands and the corresponding magnetic moments of the adatoms. For V and Ni we observe only a minor spin splitting, whereas a larger separation between the $3d$ bands can be observed for Cr, Mn, Fe, and Co. Especially for the Mn adatom an asymmetric occupation of the majority and minority channel becomes noticeable with the former being almost completely occupied and sharply localized between 2 and 3 eV below $E_F$ and the latter being located in the unoccupied region above $E_F$. In addition, one sees the rising electronic occupation of the $3d$ shell from V to Mn by means of a shift of the majority bands (minority bands in the case of Cr and V due to the AFM alignment) towards lower energies. Since the majority states are fully occupied  at the middle of the series in the case of the Mn adatom, one can see an analogous shift of minority bands as one goes from Fe to Ni. This effect is also known to appear for free $3d$ atoms.

\subsection{Electronic structure and spin polarization}
\label{sec:estructure}
In this section we analyze the electronic structure in detail for each $3d$ TM adatom on the Mn/W(110) surface within a small energy range ($\pm$1 eV) around $E_F$ which is typically accessible to STM. In this connection we first recall the most important aspects of the already investigated properties of the Co adatom ~\cite{Caffrey2014a, Serrate2010} near the Fermi level before proceeding with the other adatoms from the $3d$ TM series.  
Here, we begin our discussions with the Fe adatom focusing on the local density of states (LDOS) and its experimentally accessible spin polarization in the vacuum. Special attention will also be turned to the hybridization between the adatom $3d$ orbitals and the underlying Mn monolayer. We chose Fe as a starting point since the neighboring Co adatom in the periodic table on Mn/W(110) serves as a reference system for comparison concerning the electronic properties.

\subsubsection{Co adatom on Mn/W(110)}
\begin{figure}[hbt]
	\centering
	\includegraphics[scale=0.9,clip]{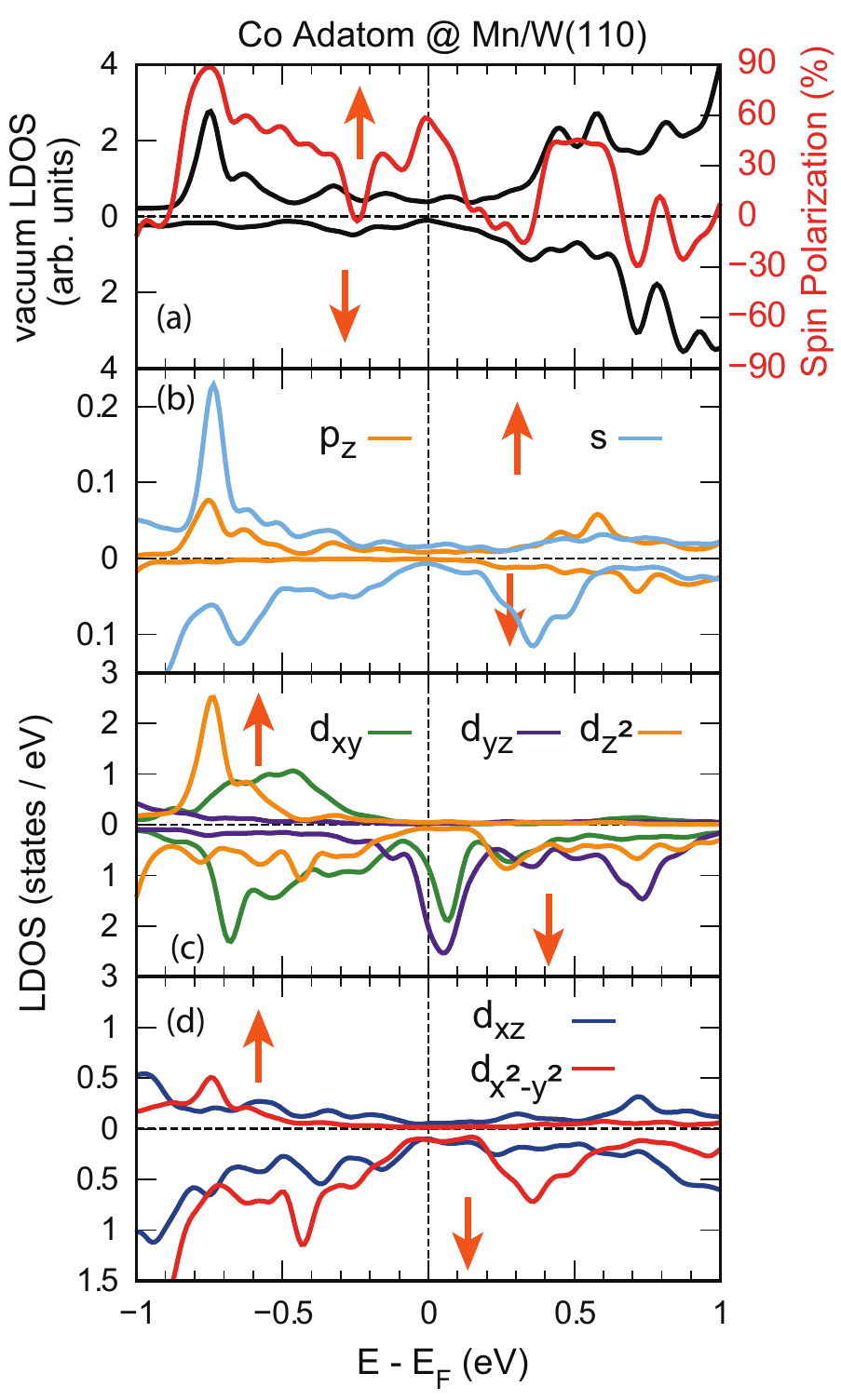}
	\caption{(a) Spin-resolved LDOS in the vacuum calculated 6 {\AA} above the Co adatom on Mn/W(110) for an out-of-plane magnetization direction, {\ie} including SOC. The spin polarization of the vacuum LDOS according to Eq. (\ref{eq:spinpolarization}) is depicted in red color. (b) Spin-resolved LDOS of the $s$ and $p_z$ states of the Co adatom. (c) and (d) orbital decomposition of the LDOS of the Co adatom in terms of the majority (up) and minority (down) states as well as the orbital symmetry of the $d$-states. The orange up and down arrows indicate majority- and minority-spin channels, respectively.  }
	\label{fig:CoAdatom_spinpol}
\end{figure}

The orbital decomposition of the spin-resolved LDOS of the Co adatom on Mn/W(110) is plotted in Figs.~\ref{fig:CoAdatom_spinpol}(b)-~\ref{fig:CoAdatom_spinpol}(d). In accordance with previous findings, $E_F$ is dominated by minority states of $d_{yz}$ and $d_{xy}$ character whose double-lobed structure can be observed in SP-STM experiments~\cite{Caffrey2014a, Serrate2010}. The majority states on the other hand are almost completely occupied and therefore quite flat in the vicinity of the $E_F$. One aspect that has not been investigated yet is the spin polarization in the vacuum which is defined as 
\begin{equation}
P(z, \epsilon)=\frac{n^{\uparrow}(z, \epsilon)-n^{\downarrow}(z, \epsilon)}{n^{\uparrow}(z, \epsilon)+n^{\downarrow}(z, \epsilon)} \times 100\%
\label{eq:spinpolarization}
\end{equation}
with $n^{\uparrow}$ ($n^{\downarrow}$) denoting the majority (minority) states with respect to the NN Mn atoms of the Mn monolayer on W(110). These quantities are shown in Fig.~\ref{fig:CoAdatom_spinpol}(a) along with the corresponding spin-resolved vacuum LDOS \footnote{This quantity can also be evaluated directly at the adatom}. While it exhibits an already large value of $+58$\% directly at the Fermi level, the largest value of $+88$\% can be found at approximately $-0.75$ eV. Smaller values of this quantity with up to $+45$\% are found in the unoccupied region at $+0.5$ eV; here also negative values of up to $-29$\% can be observed. A closer look at the orbitally resolved LDOS of the adatom reveals that due to the large spin splitting of the $3d$ states most of the high values in the spin polarization are caused by the much less split $p_z$ and $s$ states of both spin channels of the adatom [see Fig.~\ref{fig:CoAdatom_spinpol}(b)]. Apart from a hybrid $s$-$p_z$-$d_{z^2}$ state at $-0.75$ eV generating the largest positive value of $+88$\% it is mainly their peak structure which is visible in the vacuum LDOS.   
\subsubsection{Fe adatom on Mn/W(110)}
\begin{figure}[hbt]
	\centering
	\includegraphics[scale=0.9,clip]{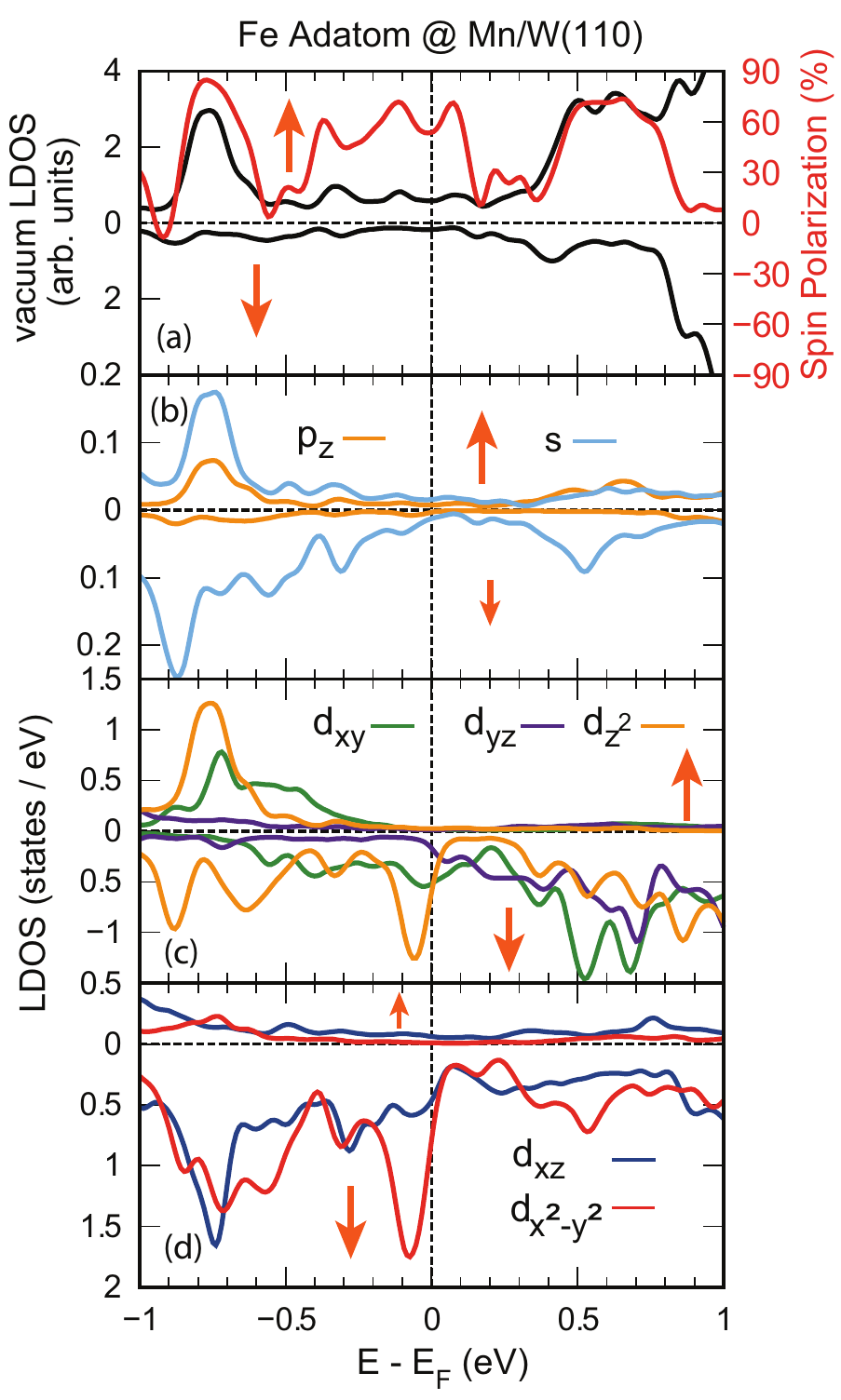}
	\caption{(a) Spin-resolved LDOS in the vacuum calculated 6 {\AA} above the Fe adatom on Mn/W(110) for an out-of-plane magnetization direction, {\ie} including SOC. The spin polarization of the vacuum LDOS according to Eq. (\ref{eq:spinpolarization}) is depicted in red color. (b) Spin-resolved LDOS of the $s$ and $p_z$ states of the Fe adatom. (c) and (d) orbital decomposition of the LDOS of the Fe adatom in terms of the majority (up) and minority (down) states as well as the orbital symmetry of the $d$-states. The orange up and down arrows indicate majority- and minority-spin channels, respectively.  }
	\label{fig:FeAdatom_spinpol}
\end{figure}

Figure~\ref{fig:FeAdatom_spinpol}(a) shows the spin-resolved LDOS calculated in the vacuum at a height of 6 {\AA} above the Fe adatom. While the minority states are quite flat in the presented energy window and do not exhibit any features, the majority states are characterized by a few clearly discernible local maxima. One of the local maxima can be observed at 0.77 eV below $E_F$, whereas a double peak structure becomes visible between 0.5 and 0.7 eV above $E_F$. A closer look at the orbitally decomposed LDOS of the adatom  [Figs.~\ref{fig:FeAdatom_spinpol}(b)-(d)] reveals that the prominent feature in the vacuum below $E_F$ stems from the majority hybrid $p_z$-$d_{z^2}$ and $s$ states which could already be observed for the Co adatom [cf. Fig.~\ref{fig:CoAdatom_spinpol}]. However, the double peak structure above $E_F$ arises due to the presence of majority $s$ and $p_z$ states. Orbitals of this symmetry were expected to contribute in large part to the LDOS in the vacuum due to their strong delocalization. In contrast, dominant localized minority peaks with $d_{x^2 - y^2}$ and $d_{xy}$ character which are also present above $E_F$ [Fig.~\ref{fig:FeAdatom_spinpol}(d)] are not visible in the vacuum since their lobes lie in-plane. 

The above mentioned striking features of the vacuum LDOS give rise to large positive values of the spin polarization as well which is shown on a second axis in Fig.~\ref{fig:FeAdatom_spinpol}(a). At the respective positions of the local maxima, the spin polarization takes on values ranging from 74\% up to 84\%, while the effect is slightly smaller at $E_F$ with a value of 55\%. However, this value is still nearly three times higher than the one of 20\% which has recently been reported for the clean Mn/W(110) surface~\cite{Hauptmann2019}. As already shown in Table~\ref{tab:table2}, a Mn atom in the monolayer on the W(110) substrate exhibits a large spin moment of 3.41 $\mu_B$ thereby causing a large exchange splitting of the $3d$ bands which are shifted away from $E_F$. Instead the less spin polarized $p_z$ orbitals dominate at $E_F$~\cite{Hauptmann2019}. In the case of a deposited Fe adatom the much higher spin polarization can be explained by the $p_z$ and additionally $s$ states at $E_F$ as well, especially the ones with majority character (see Fig.~\ref{fig:FeAdatom_spinpol}(b)).

Comparing further the electronic structure of the Co adatom on Mn/W(110) ~\cite{Serrate2010, Caffrey2014a, Haldar2018} with the one of a Fe adatom, one sees that the LDOS of the latter is dominated by minority states with $d_{z^2}$ and $d_{x^{2}-y^{2}}$ states at $E_F$. The minority states with $d_{yz}$ and $d_{xy}$ character which were found at $E_F$ for the Co adatom are shifted towards higher energies of 0.5 to 0.7 eV above $E_F$ for the Fe adatom. This is due to the larger spin splitting and 
consequent
different magnetic moment on the one hand and due to one missing electron in the $3d$ orbitals on the other hand (see Fig.~\ref{fig:FeAdatom_spinpol}(c)). In contrast to the minority states, the majority $3d$ as well as $p$ bands of the Fe adatom are not affected by this energetic shift (see Fig.~\ref{fig:3ddos}).  

\begin{figure}[htbp]
	\centering
	\includegraphics[scale=0.9,clip]{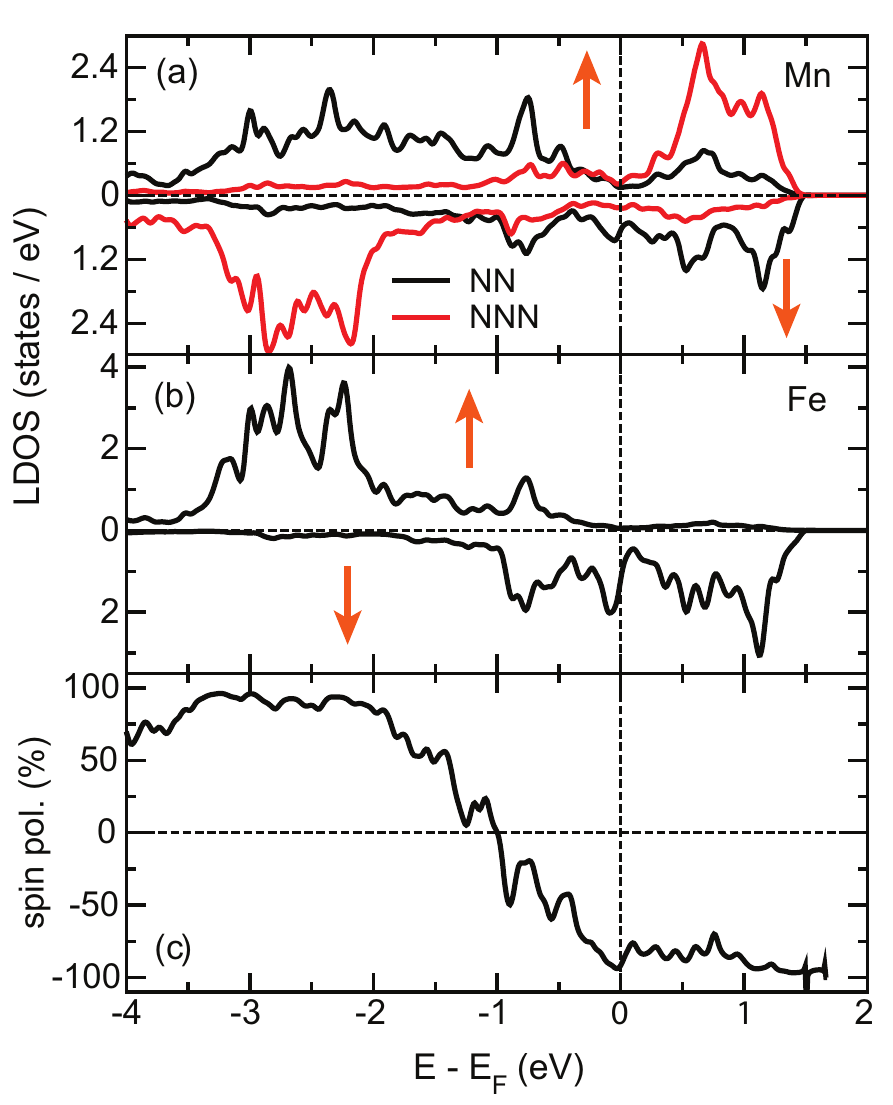}
	\caption{(a) Spin-resolved LDOS of the nearest neighbor (NN) and next-nearest neighbor (NNN) Mn atoms of the Fe adatom in the Mn monolayer on the W(110) surface. (b) Spin-resolved LDOS of the Fe adatom (sum of $3d$ orbitals). (c) Spin polarization of the Fe adatom as in Eq. (\ref{eq:spinpolarization}) using the spin-resolved LDOS at the Fe atom. The orange up and down arrows indicate majority- and minority-spin channels, respectively.}
	\label{fig:Fe_Atom_spinpol}
\end{figure}
In the next step we take a closer look at the hybridization between the Fe adatom and the substrate. This is firstly exemplified by comparing the spin-resolved LDOS of the adsorbate and its neighboring Mn atoms shown in Figs.~\ref{fig:Fe_Atom_spinpol}(a) and~\ref{fig:Fe_Atom_spinpol}(b). As already indicated before, hybridization between different states can be recognized by peaks in the LDOS being located at the same energy position, e.g. at $-0.75$ eV between the dominant majority $d_{z^2}$ state of the Fe adatom and the majority $3d$ states of the NN Mn atoms on the surface (the same behavior is observed for the respective minority states at this energetic position). Further evidence of such interaction can be found just below $E_F$ for both minority Fe and NN Mn states and at $-3$ eV in the majority spin channel of the adatom and the substrate.
The large spin splitting of the Fe adatom becomes clearly visible resulting in a magnetic moment of +2.79 {\uB}. As mentioned earlier, the majority bands are nearly occupied, whereas the minority states predominating at $E_F$ are only half full. 

This aspect can also be seen in the spin polarization calculated directly at the Fe adatom [Fig.~\ref{fig:Fe_Atom_spinpol}(c)]. Consistent with the LDOS, this quantity changes sign at $-1$ eV dropping from maximum values of +96\% at $-3$ eV to a local minimum of $-93$\% directly at $E_F$. Since these energetic positions are matching quite well with the ones mentioned above with respect to the hybridization between the adatom and the surface, one can conclude that the high spin polarization of the Fe adatom is due to spin splitting of $d$-bands.  It should also be pointed out here that the large negative spin polarization in the vicinity of the $E_F$ of the Fe adatom is a result of its minority $d$ states covering the contributions of $s$ and $p$ orbitals (not shown). However, as pointed out earlier ~\cite{Ferriani2010}, the delocalized majority $p_z$ states generate large positive values of up to 55\% in the vacuum close to the $E_F$ instead (cf. Fig. 4(a)).  

\begin{figure}[htbp]
	\centering
	\includegraphics[scale=0.8,clip]{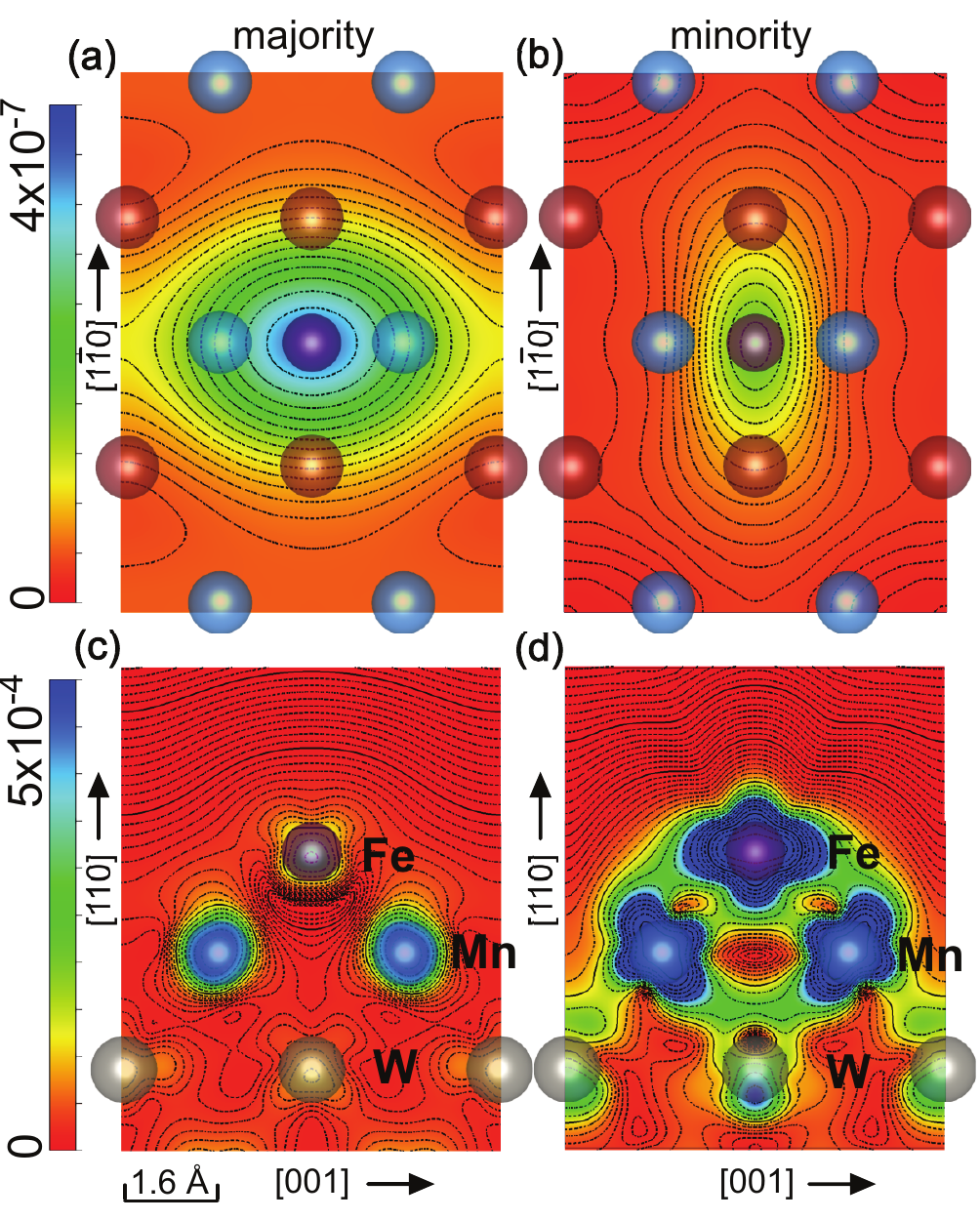}
	\caption{(a, b): Spin-resolved partial charge density calculated at a height of 3 {\AA} above the Fe adatom on Mn/W(110) in the energy interval [$E_F-0.064$, $E_F-0.044$ eV]. (c, d): Side view of the respective partial charge densities shown above represented as cross-sectional plots through the Fe adatom parallel to the $[001]$ direction. The zero point on the height scale is defined by the position of the Mn monolayer. 
	Both
	majority and minority spin channels
	are plotted on the same color scale in order to ensure comparability of the two spin channels. }
	\label{fig:chargedensity_Fe}
\end{figure}
In order to visualize the hybridization between the Fe adatom and the Mn atoms of the Mn monolayer on W(110) more clearly we calculated the spin-resolved partial charge density in a small energy range below $E_F$ where dominant peaks in the LDOS can be observed. This is the case for the minority $d_{x^{2}-y^{2}}$ and $d_{z^2}$ orbitals of the Fe adatom at approximately $E_F-0.05$ eV [see Figs.~\ref{fig:FeAdatom_spinpol}(c) and~\ref{fig:FeAdatom_spinpol} (d)]. Figure~\ref{fig:chargedensity_Fe} shows both the top view at a height of 3 {\AA} above the Fe adatom in the vacuum as well as the side view parallel to the $[001]$ direction of the unit cell at the aforementioned energy. For both spin channels one can identify rotationally symmetric orbitals from the top which are elongated along the $[001]$ direction for the case of the majority states and squeezed for the case of the minority states, respectively [Figs.~\ref{fig:chargedensity_Fe}(a) and~\ref{fig:chargedensity_Fe}(b)]. The corresponding cross-sectional plots reveal that the majority spin channel of the Fe adatom exhibits a state composed of $d_{xz}$, $p_z$, and $s$ orbitals at this energy [Fig.~\ref{fig:chargedensity_Fe}(c)]. However, the double-lobed structure of the low-intensity $d_{xz}$ state is completely covered by the rotationally symmetric $s$ orbitals and $p_z$ orbitals in the vacuum. The minority spin channel [Fig.~\ref{fig:chargedensity_Fe} (d)] shows a mixed state consisting of the above-mentioned $d_{z^2}$ and $d_{x^{2}-y^{2}}$ orbitals at this energy which strongly hybridizes with the $d_{z^2}$ states of the NN Mn atoms. Their orbitals are not aligned parallel to the surface normal anymore, but tilted towards the adatom orbitals instead [Fig.~\ref{fig:chargedensity_Fe}(d)]. Besides, one can observe an accumulation of the charge density at the interface between the substrate and the adatom (see green shaded area in Fig.~\ref{fig:chargedensity_Fe}(d)).  

In comparison for the Co adatom on Mn/W(110) \cite{Caffrey2014a, Serrate2010, Haldar2018}, it has been shown that the majority spin channel is characterized by a composite state of rotationally symmetric $s$, $p_z$ and $d_{z^2}$ orbitals, whereas the minority channel is dominated by a double-lobed $d_{yz}$ orbital in the vicinity of $E_F$. Therefore, using SP-STM it is possible to display both spin channels independently and hence read out the spin orientation of the Co adatom by a magnetic tip \cite{Serrate2010, Serrate2016}. However, our calculations reveal that the same cannot be realized for the Fe adatom adjacent in the $3d$ row, at least by imaging at typically small bias voltages, due to the rotational symmetry of the orbitals which are visible in the vacuum. 

\subsubsection{Ni adatom on Mn/W(110)}
\begin{figure}[hbtb]
	\centering
	\includegraphics[scale=0.9,clip]{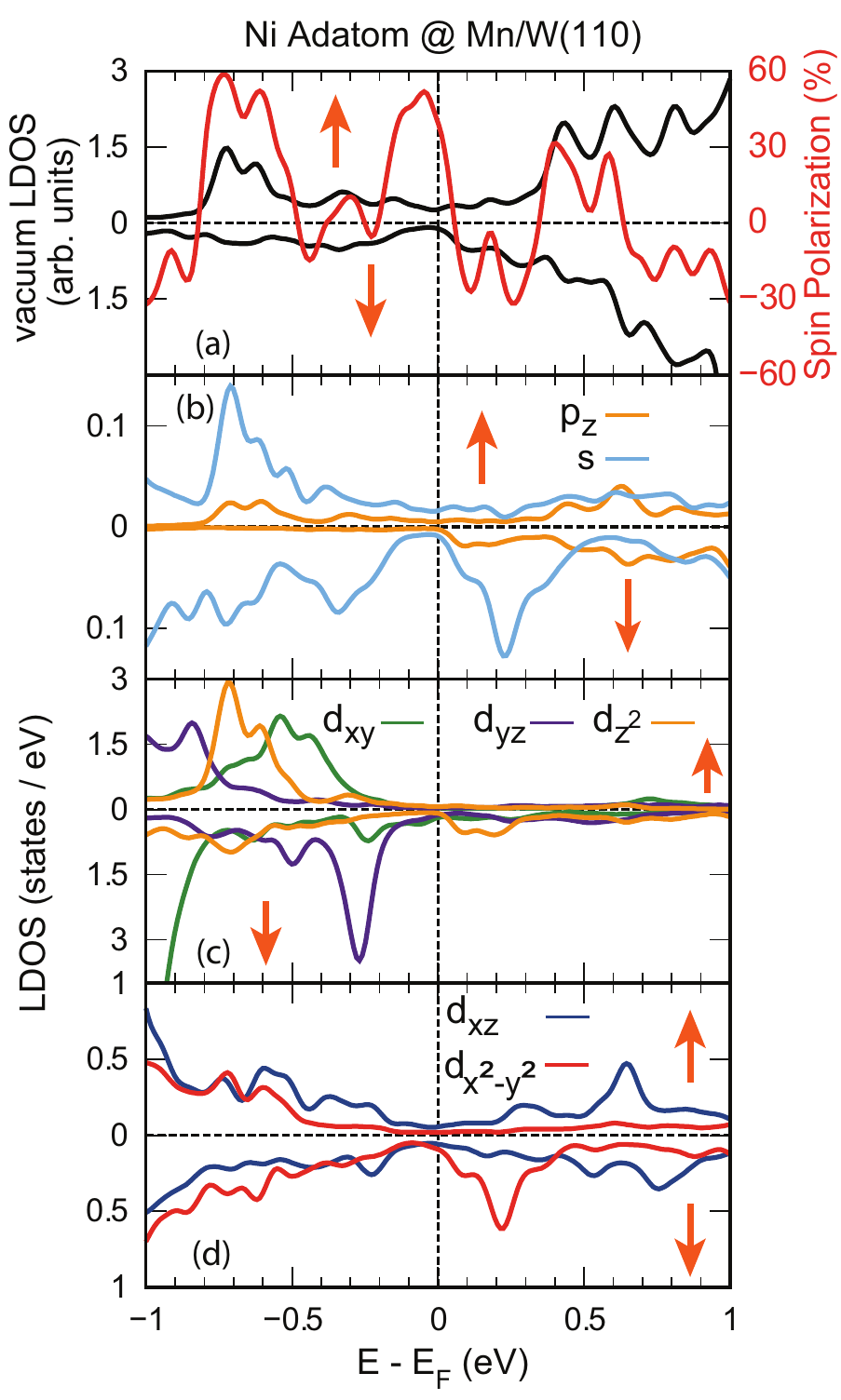}
	\caption{(a) Spin-resolved LDOS in the vacuum calculated 6 {\AA} above the Ni adatom on Mn/W(110) for an out-of-plane magnetization direction, {\ie} including SOC. The spin polarization of the vacuum LDOS according to Eq. (\ref{eq:spinpolarization}) is depicted in red color. (b) Spin-resolved LDOS of the $s$ and $p_z$ states of the Ni adatom. (c) and (d) orbital decomposition of the LDOS of the Ni adatom in terms of the majority (up) and minority (down) states as well as the orbital symmetry of the $d$-states. The orange up and down arrows indicate majority- and minority-spin channels, respectively.   }
	\label{fig:NiAdatom_spinpol}
\end{figure}
Next we consider the electronic structure of a single magnetic Ni adatom on the Mn/W(110) surface and compare it with the previously presented density of states of the Co adatom on the Mn monolayer on W(110). Ni possesses one more electron in its $3d$ shell than
Co and carries a much smaller magnetic moment of +0.39 {\uB} upon adsorption on the Mn monolayer. This leads to a reduced spin splitting of the states compared to the adsorbed Co adatom,
which has a magnetic moment of +1.61 {\uB}. These differences with respect to the occupation and magnetic moment can be clearly seen in the LDOS of the Ni $s$, $p$ and $d$ states in Figs.~\ref{fig:NiAdatom_spinpol}(b)-\ref{fig:NiAdatom_spinpol}(d). 

The minority states are shifted towards lower energies which can partly be explained by the above mentioned increasing 
occupation of electrons and partly by the small spin-splitting. As a result both majority and minority $d$ states are almost entirely occupied and $E_F$ is dominated by $p_z$ and $s$ states. For instance, the minority state with $d_{yz}$ symmetry which was found at $E_F$ of Co is shifted towards lower energies by 0.3 eV in the case of Ni [Fig.~\ref{fig:NiAdatom_spinpol}(c)]. The same holds true for a peak of minority $d_{x^{2}-y^{2}}$ orbitals at 0.21 eV which is located at a higher energy for Co. The majority states below  $E_F$ on the other hand are likewise dominated by the large peak of the hybrid $s$- $p_z$-$d_{z^2}$ state around $-0.71$ eV which experiences a splitting into two local maxima in the case of the Ni adatom. 

The above mentioned two peaks also represent the most noticeable feature of the vacuum LDOS in the occupied region causing a maximum value in the spin polarization of +59\% as it can be seen in Fig.~\ref{fig:NiAdatom_spinpol}(a). Directly at $E_F$ the spin polarization is in the same order of magnitude and can be attributed to $p_z$ and $s$ states only similar to the case of Co and Fe. As for the Co and Fe adatom, the unoccupied energy range in the vacuum is dominated by states of $s$ and $p_z$ symmetry which leads to positive and negative values in the spin polarization. For example, this can be observed at approximately 0.26 eV where a large component of minority bands exceeds the majority curve thereby generating a spin polarization of up to $-31$\% according to Eq.~(\ref{eq:spinpolarization}). Further it should be noted that between 0.4 eV and 0.6 eV above $E_F$ the same two majority $p_z$ peaks can be observed which are also present in the case of the Co adatom on Mn/W(110).  
     
\subsubsection{Mn adatom on Mn/W(110)}
\label{subsubsec:Mn_Adatom}
\begin{figure}[hbtb]
	\centering
	\includegraphics[scale=0.9,clip]{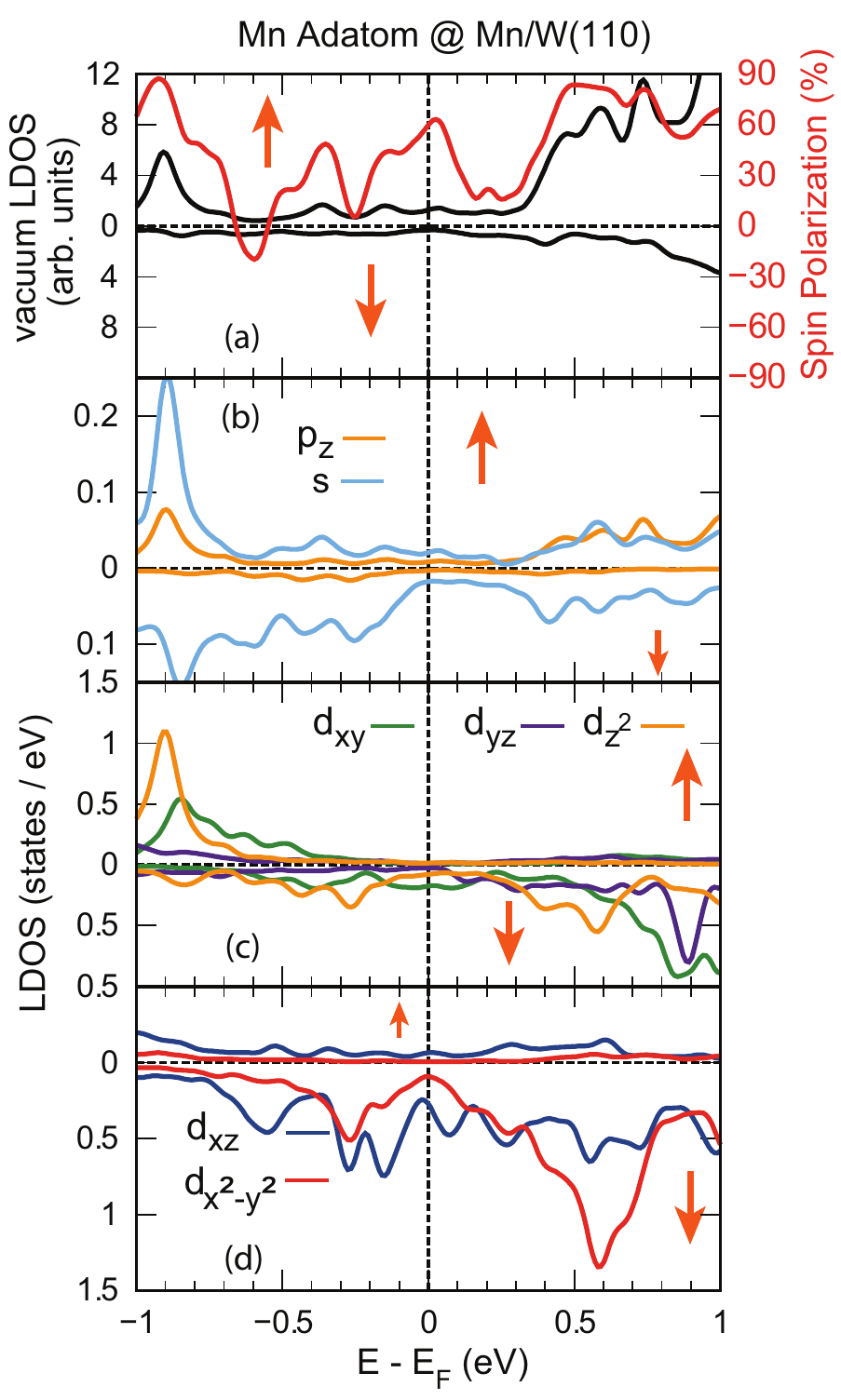}
	\caption{(a) Spin-resolved LDOS in the vacuum calculated 6 {\AA} above the Mn adatom on Mn/W(110) for an out-of-plane magnetization direction, {\ie} including SOC. The spin polarization of the vacuum LDOS according to Eq. (\ref{eq:spinpolarization}) is depicted in red color. (b) Spin-resolved LDOS of the $s$ and $p_z$ states of the Mn adatom. (c) and (d) orbital decomposition of the LDOS of the Mn adatom in terms of the majority (up) and minority (down) states as well as the orbital symmetry of the $d$-states. The orange up and down arrows indicate majority- and minority-spin channels, respectively.  }
	\label{fig:MnAdatom_spinpol}
\end{figure}
Here we investigate the local density of states of a single magnetic Mn adatom adsorbed on the Mn monolayer on W(110). In comparison with the neighboring Fe and the reference Co adatom, the $3d$ states of the Mn adatom are characterized by an even larger spin splitting on the one hand and less electronic occupation of the $d$ shell on the other hand.

Fig.~\ref{fig:MnAdatom_spinpol}(a) shows the spin resolved vacuum LDOS calculated at 6 {\AA} above the Mn adatom. As for the Fe adatom, it is mainly dominated by peaks of majority states while the minority LDOS is flat and does not exhibit any special features. The orbitally decomposed LDOS of Mn [Figs.~\ref{fig:MnAdatom_spinpol}(b)-\ref{fig:MnAdatom_spinpol}(d)] shows that apart from the majority hybrid $s$-$p_z$-$d_{z^2}$ state which is also present at $-0.92$ eV for this adatom the vacuum LDOS is mainly composed of states with $s$ and $p_z$ character again. It is also clear that the energetic positions of the unoccupied majority $s$ and $p_z$ bands are independent of the chosen $3d$ TM adatom since there are no $d$ states available for this spin character in that energy range. Similar as for Fe adatom on Mn/W(110), the spin polarization of the Mn adatom takes large positive values caused by $s$ and $p_z$ states and ranging from 87\% at $-0.92$ eV due to the presence of the majority hybrid $s$-$p_z$-$d_{z^2}$ state to 63\% close to $E_F$.      

In addition, it is apparent by means of the orbitally decomposed LDOS of the Mn adatom that the majority $3d$ states are almost entirely occupied. 
As discussed in the section~\ref{subsec:str_mag}, these states are sharply localized in the energy interval between $-3$ eV and $-2$ eV (see Fig.~\ref{fig:3ddos}).
In contrast to the Fe adatom, here $E_F$ is predominated by minority states with $d_{xz}$ character. The shift of the bands due to the different magnetic moment and occupation is also clearly visible in this case. The peaks with minority $d_{z^2}$ and $d_{x^{2}-y^{2}}$ states, which are observed below $E_F$ for the Fe adatom, are now moved by 0.6 eV towards the unoccupied region. The same applies to the minority $d_{yz}$ state dominating the $E_F$ of the reference adatom Co; it is located at 0.9 eV above $E_F$ in the case of Mn [see Fig.~\ref{fig:MnAdatom_spinpol}(c)].  

\subsubsection{Cr adatom on Mn/W(110)}
\begin{figure}[htbp]
	\centering
	\includegraphics[scale=0.9,clip]{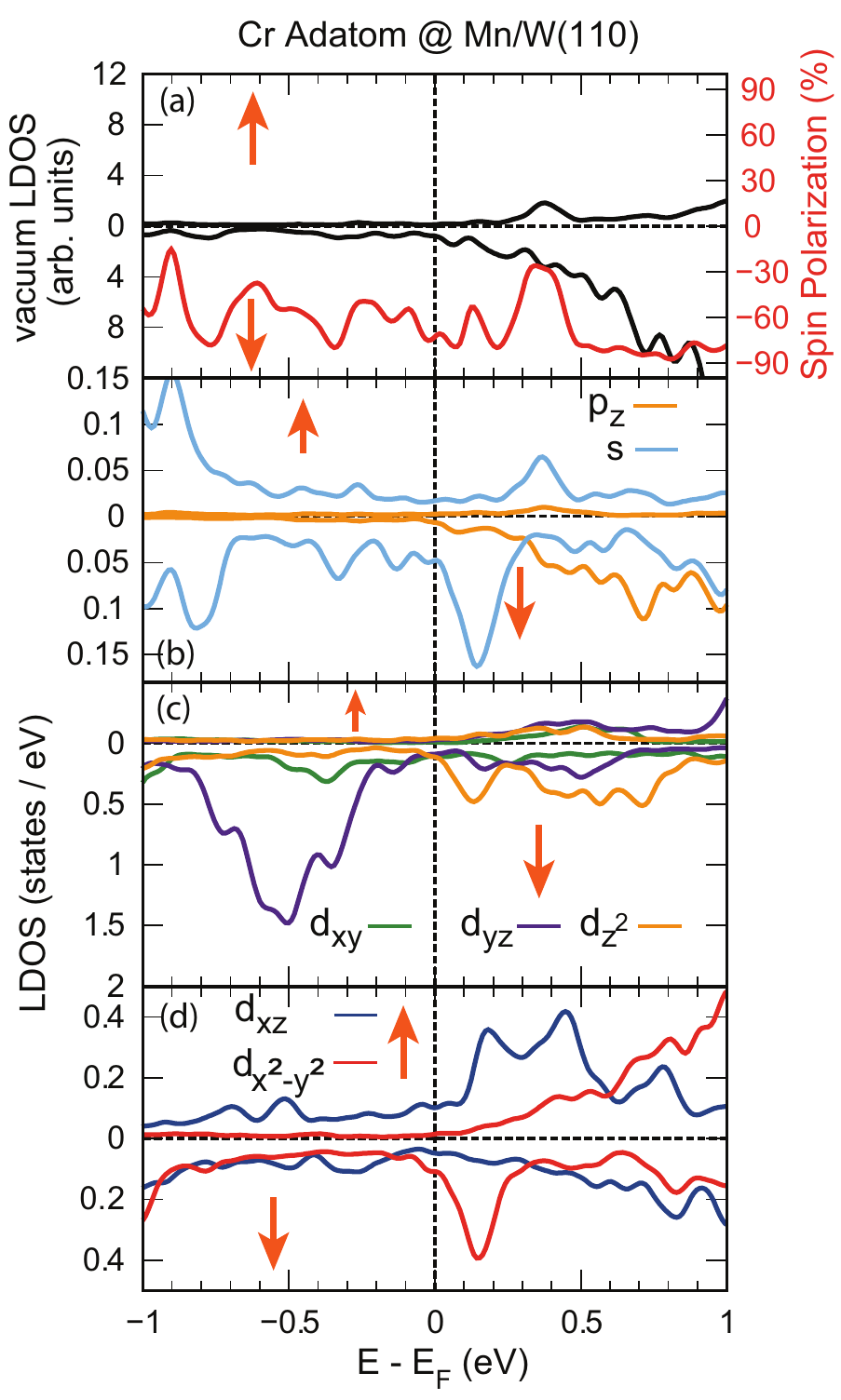}
	\caption{(a) Spin-resolved LDOS in the vacuum calculated 6 {\AA} above the Cr adatom on Mn/W(110) for an out-of-plane magnetization direction, {\ie} including SOC. The spin polarization of the vacuum LDOS according to Eq. (\ref{eq:spinpolarization}) is depicted in red color. Note that majority and minority spin channels are defined with respect to the NN Mn atoms to which Cr couples antiferromagnetically (cf. Fig.~\ref{fig:3ddos}). (b) Spin-resolved LDOS of the $s$ and $p_z$ states of the Cr adatom. (c) and (d) orbital decomposition of the LDOS of the Cr adatom in terms of the majority (up) and minority (down) states as well as the orbital symmetry of the $d$-states. The orange up and down arrows indicate majority- and minority-spin channels, respectively. }
	\label{fig:CrAdatom_spinpol}
\end{figure}
Now we consider the electronic structure of a Cr adatom on Mn/W(110). In contrast to the reference Co adatom, Cr carries a much larger magnetic moment of $-3.69$ {\uB} thereby realizing an AFM alignment with its NN Mn atoms of the Mn monolayer. The higher magnetic moment is accompanied by an increased spin splitting of the $3d$ states, {\ie} they are expected to move further apart. As in the case of Mn, both the larger spin splitting of the states and the reduced occupation of the majority bands compared to the Co adatom become evident in the LDOS of the Cr adatom shown in Figs.~\ref{fig:CrAdatom_spinpol}(c) and~\ref{fig:CrAdatom_spinpol}(d). The minority bands are densely populated by states of $d_{x^{2}-y^{2}}$, $d_{z^2}$ and a prominent peak of $d_{yz}$ symmetry for the presented energy range in the case of the Cr adatom. This behavior is different from the neighboring Mn adatom, for which the similar majority states (since it is coupled ferromagnetically with the NN Mn atoms) are already shifted further below $E_F$ due to the increased electronic occupation. The minority $d_{xz}$ bands dominating the Fermi level of Mn [cf. Figs.~\ref{fig:MnAdatom_spinpol}(d)] can be found at higher energies between 0.2 and 0.5 eV in the majority channel in the case of Cr now. In exactly the same way as for the previous investigated $3d$ TM adatoms on Mn/W(110), the structure of the $p_z$ and $s$ states emerges in the vacuum LDOS at a height of 6 {\AA} above the Cr adatom leading to large negative values of $-73\%$ (due to the AFM coupling) for the spin polarization in the vicinity of $E_F$ [Fig.~\ref{fig:CrAdatom_spinpol}(a)]. The most prominent feature of the vacuum density of states of the previous investigated adatoms - the dominant majority hybrid $s$-$p_z$-$d_{z^2}$ state - is missing in the minority spin channel for the Cr adatom as it can be seen from Figs. ~\ref{fig:CrAdatom_spinpol}(a) -~\ref{fig:CrAdatom_spinpol}(c). Though there is still a small peak of $d_{z^2}$ character at $\approx$ 0.1 eV above $E_F$, it does not interact with the $p_z$ states anymore, but rather with a minority $d_{x^{2}-y^{2}}$ state at the same energy.

\subsubsection{V adatom on Mn/W(110)}
\begin{figure}[hbt]
	\centering
	\includegraphics[scale=0.9,clip]{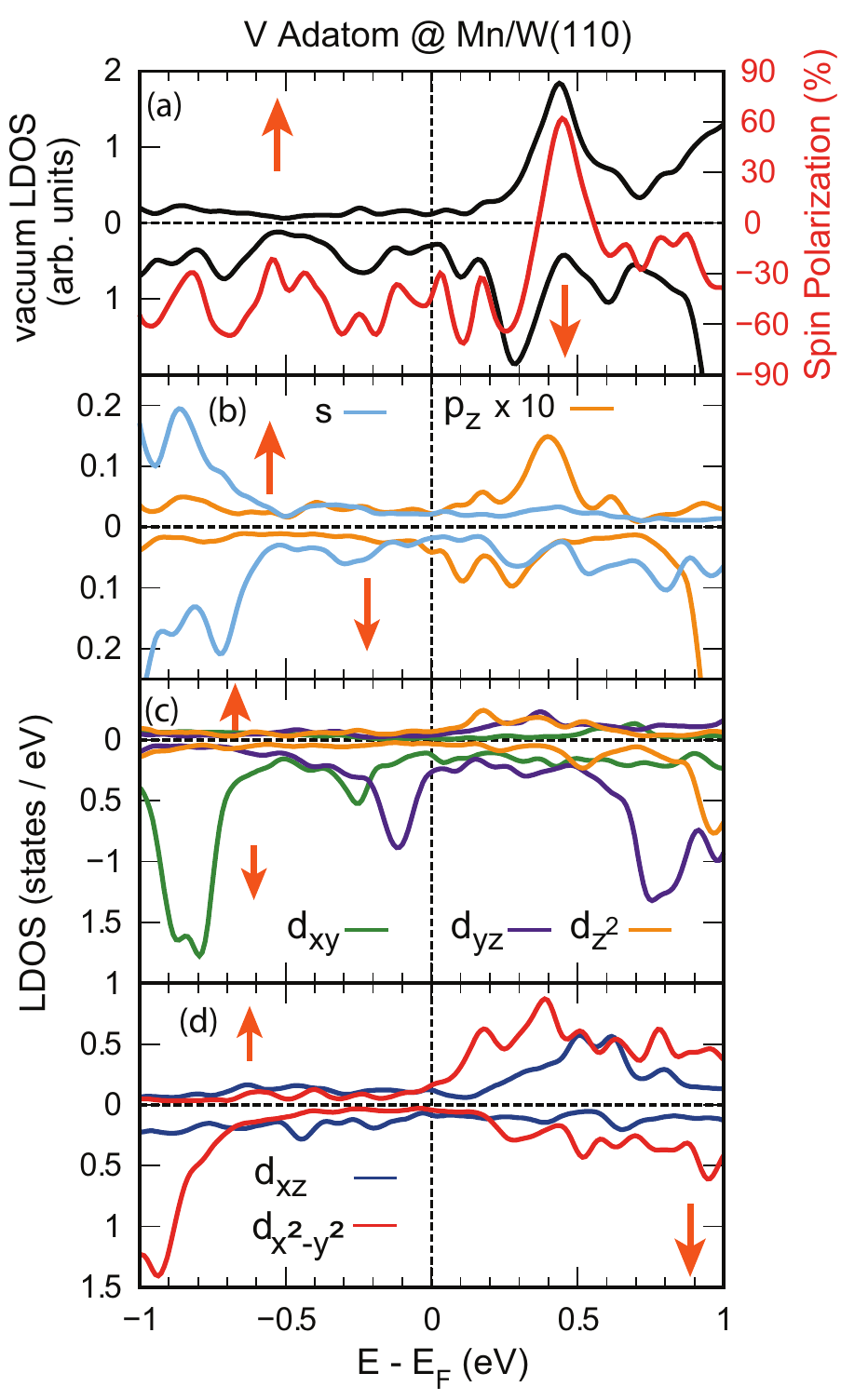}
	\caption{(a) Spin-resolved LDOS in the vacuum calculated 6 {\AA} above the V adatom on Mn/W(110) for an out-of-plane magnetization direction, {\ie} including SOC. The spin polarization of the vacuum LDOS according to Eq. (\ref{eq:spinpolarization}) is depicted in red color. Note that majority and minority spin channels are defined with respect to the NN Mn atoms to which V couples antiferromagnetically (cf. Fig.~\ref{fig:3ddos}) (b) Spin-resolved LDOS of the $s$ and $p_z$ states of the V adatom. (c) and (d) orbital decomposition of the LDOS of the V adatom in terms of the majority (up) and minority (down) states as well as the orbital symmetry of the $d$-states. The orange up and down arrows indicate majority- and minority-spin channels, respectively.  }
	\label{fig:VAdatom_spinpol}
\end{figure}
The last adatom of the $3d$ series that we have investigated is V. In bulk it is non-magnetic but, as we have seen in section~\ref{subsec:str_mag}, it develops a magnetic moment of $-1.54$ {\uB} upon adsorption on the Mn/W(110) surface. Hence a spin splitting between its $3d$ states can be observed due to hybridization with the substrate. The spin resolved LDOS calculated in the vacuum for this adatom is plotted in Fig.~\ref{fig:VAdatom_spinpol}(a). It is mainly dominated by $p_z$ and $s$ states just below and above $E_F$ as evident from a comparison with Fig.~\ref{fig:VAdatom_spinpol}(b). Here contributions of both spin channels become visible resulting in numerous oscillations in the curve of the spin polarization. While this quantity is very small directly at $E_F$ with values up to $-38\%$, it also exhibits a large positive value of 62\% at 0.45 eV above the Fermi level and large negative values of up to $-66$\% in the occupied regions.

In contrast to the neighboring Cr adatom, the Fermi level of the V adatom is dominated by a prominent peak of minority $d_{yz}$ states (see Fig.~\ref{fig:VAdatom_spinpol}(c)), which can be found outside of the presented energy region at $\approx$ 1 eV in the case of Cr. Since this state decays quickly in the vacuum, it is not visible at a height of 6 {\AA} in the vacuum above the V adatom. The striking $d_{yz}$ minority peak dominating the LDOS of Cr at $-0.5$ eV (see Fig.~\ref{fig:CrAdatom_spinpol}(c)) is shifted to a higher energetic position of 0.8 eV in the case of V which can be attributed to both smaller spin splitting and less electronic occupation.

\begin{figure}[htb]
	\centering
	\includegraphics[scale=0.8,clip]{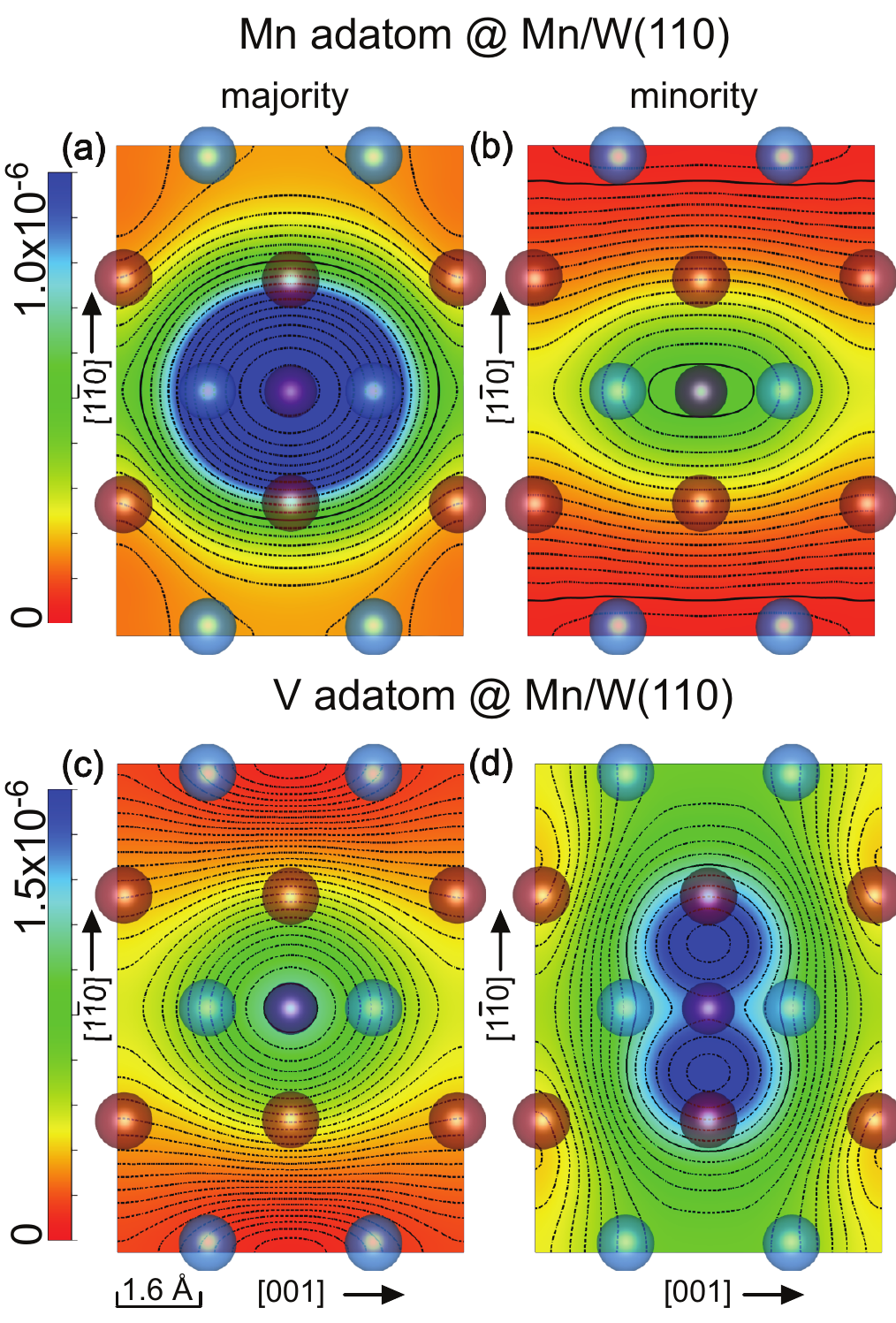}
	\caption{(a, b) Spin-resolved partial charge density calculated at a height of 3 {\AA} above the Mn adatom on Mn/W(110) in the energy interval [$E_F$, $E_F+0.1$ eV]. (c, d) Spin-resolved partial charge density calculated at a height of 3 {\AA} above the V adatom on Mn/W(110) in the energy interval [$E_F-0.15$, $E_F-0.05$ eV]. 
	Both majority and minority spin channels are plotted on the same color scale in order to ensure comparability of the two spin channels.
	 }
	\label{fig:chargedensity_Mn_V}
\end{figure}
Now we turn our focus again to the question whether both spin channels of a single adsorbed $3d$ TM adatom on the Mn/W(110) substrate can be identified independently in a SP-STM experiment. Our calculations of the respective charge densities close to $E_F$ have shown that just like for the Fe adatom this is not possible for Ni, Mn and Cr either. Even though double-lobed $d_{xz}$ orbitals dominate the spin channels of the Mn and the Cr adatom near $E_F$ [see Figs.~\ref{fig:MnAdatom_spinpol}(d) and ~\ref{fig:CrAdatom_spinpol}(d)], they remain hidden in the vacuum due to overlap with the $s$ and the $d_{z^2}$ states. This aspect is exemplified for the Mn adatom in Figs.~\ref{fig:chargedensity_Mn_V}(a) and~\ref{fig:chargedensity_Mn_V}(b) in which the top view at a height of 3 {\AA} in the vacuum shows spherical orbitals for both spin channels.

However, for the V adatom [Fig.~\ref{fig:chargedensity_Mn_V} (c) and (d)] one can observe the spherical and double-lobed structure found in the SP-STM at small bias voltages for the majority and the minority spin channel, respectively, seen in the case of a Co adatom on Mn/W(110)~\cite{Serrate2010,Caffrey2014a,Haldar2018}. 
As already shown by means of the orbitally decomposed LDOS in Fig.~\ref{fig:VAdatom_spinpol}(c), the minority states of this adatom are characterized by a prominent peak of $d_{yz}$ orbitals just below $E_F$. A slice of the charge density at its respective energetic position at a height of 3 {\AA} above the unit cell illustrates the characteristic double-lobed structure along the $[1\overline{1}0]$ direction in the minority spin channel, while a rotationally symmetric orbital becomes visible for the majority states (see Figs.~\ref{fig:chargedensity_Mn_V}(c) and~\ref{fig:chargedensity_Mn_V}(d)). Hence, in SP-STM experiments with a magnetic STM tip the spin orientation of the V adatom could likely be similarly explicitly determined by means of the orbital shape.

\subsection{TAMR of $3d$ TM adatoms on Mn/W(110)}
\label{sec2:tamr_results}
In this section we discuss the TAMR effect driven by the adsorption of $3d$ TM adatoms on Mn/W(110). As mentioned previously in section~\ref{sec:estructure}, we discuss the TAMR effect within a small energy range ($\pm$1 eV) around $E_F$ which is typically accessible to STM. We have chosen the single magnetic Fe adatom and Mn adatom with significant magnetic moments to demonstrate this effect in detail.

The experimentally observed TAMR is defined as $[(\mathrm{d}I/\mathrm{d}V)_{\perp}-(\mathrm{d}I/\mathrm{d}V)_{\parallel}]/(\mathrm{d}I/\mathrm{d}V)_{\perp}$, where $(\mathrm{d}I/\mathrm{d}V)$ is the differential conductance and $\perp$ indicates a magnetization direction perpendicular to the surface and $\parallel$ indicates a magnetization direction parallel to the surface. 
Within the framework of the Tersoff-Hamann model~\cite{Tersoff1983,Tersoff1985}, using the spectroscopic mode of an STM,
the differential conductance $(\mathrm{d}I/\mathrm{d}V)$ is directly proportional to the LDOS at the tip position in the vacuum above the surface~\cite{Wortmann2001}. By measuring the LDOS at a specific height $z$ in the vacuum above an adatom with two different magnetization directions - perpendicular to the surface ($n_{\perp}(z,\epsilon)$) and in-plane along the $[1\overline{1}0]$ direction ($n_\parallel(z,\epsilon)$) - the value of the TAMR is calculated as:
\begin{align}
\mathrm{TAMR} &= \frac{n_\perp(z, \epsilon)-n_\parallel(z, \epsilon)}{n_\perp(z, \epsilon)} \times 100\%\; .
\label{eq:TAMR}
\end{align}

\subsubsection{TAMR of Fe adatom on Mn/W(110)}
\begin{figure}[htbp]
	\centering
	\includegraphics[scale=0.9,clip]{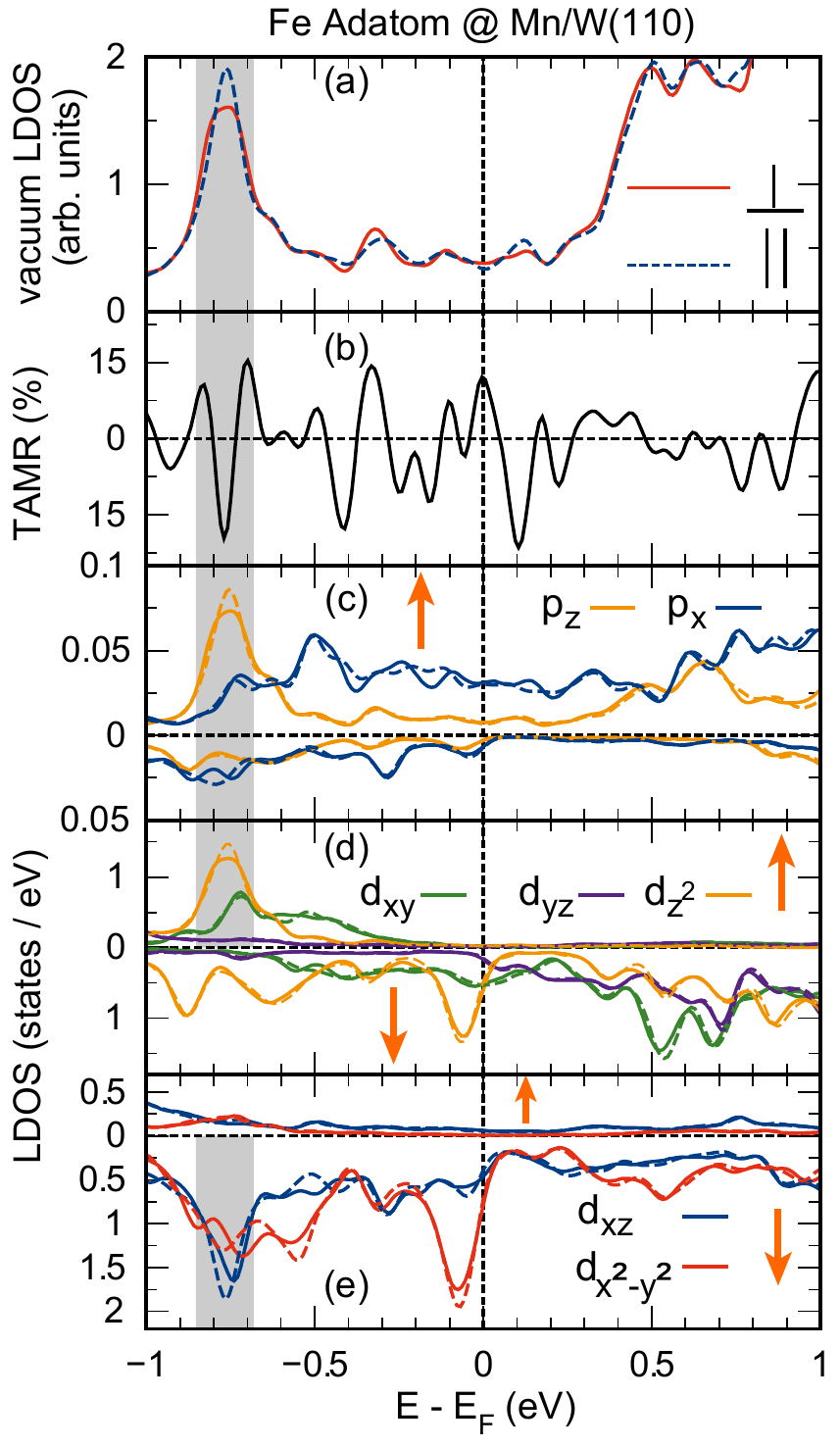}
	\caption{(a) LDOS in the vacuum calculated 6 {\AA} above the Fe adatom on Mn/W(110) for out-of-plane (solid red line) and in-plane (parallel to the $[1\overline{1}0]$ direction) magnetization directions (dashed blue line). (b) TAMR effect obtained from the data presented in (a) according to Eq.~(\ref{eq:TAMR}). (c), (d) and (e) orbital decomposition of the LDOS of the Fe adatom in terms of the majority (up) and minority (down) states as well as the orbital symmetry of the $p$- and $d$-states. Solid (dashed) lines correspond to the magnetization direction perpendicular (parallel) to the surface plane. The orange up and down arrows indicate majority- and minority-spin channels, respectively. Prominent peaks of the LDOS which will be investigated by means of a simple SOC model in section \ref{sec:tamr_model_result} are highlighted by shaded areas.}
	\label{fig:FeAdatom}
\end{figure}
Figure ~\ref{fig:FeAdatom}(a) shows the spin-averaged LDOS calculated in the vacuum at a height of $z=6${\AA} above the Fe adatom with two different magnetization directions: out-of-plane (perpendicular to the surface ($n_\perp(z, \epsilon)$)) and in-plane (along to the $[1\overline{1}0]$ direction ($n_\parallel(z, \epsilon)$))~\footnote{It is to be noted that in our calculations, the magnetization directions are applied to the complete system consisting of the adsorbed adatom and the underlying Mn/W(110) surface.}. 
Although both curves look very similar, a closer inspection reveals several differences. The most prominent difference can be observed around $-0.75$ eV. 
Here, the in-plane magnetization component is dominant over the out-of-plane component leading to a negative value of the TAMR of $-20$\% (see Fig.~\ref{fig:FeAdatom}(b)) according to Eq.~(\ref{eq:TAMR}). 
More differences are observed in the energy range between $-0.41$~eV and $-0.33$~eV leading to oscillatory TAMR values ranging from $-18$\% up to +14\%. Around 
+0.1~eV, the anisotropy of the vacuum LDOS exhibits a local minimum of $-20$\% again.
 
As already pointed out in section \ref{sec:estructure} the largest contribution to the LDOS in the vacuum comes from $d_{z^2}$ orbitals besides $s$ and $p_z$ states due to their double-lobed structure along the surface normal. Hence, the largest part of the TAMR is created due to a modification of those states from the Fe adatom, in particular from the $3d$ orbitals due to the nature of the element, but also from a modification of the $3p$ states as we will show below. In order to determine these orbital contributions,
the orbitally decomposed LDOS of the Fe 
$3p$ and $3d$  states is shown in Figs.~\ref{fig:FeAdatom}(c)-\ref{fig:FeAdatom}(e) for both magnetization orientations. It is evident that the large peak of the vacuum LDOS at $-0.75$ eV which is linked to a negative value of the TAMR corresponds to the already discussed majority $d_{z^2}$ state. At the same energy position a minority $d_{xz}$ state with discernible changes between $n_\perp$ and $n_\parallel$ is also found [Fig.~\ref{fig:FeAdatom}(e)]. 
This observation hints at a spin-orbit induced hybridization between the two states generating the TAMR. From Fig.~\ref{fig:FeAdatom}(c) it becomes also clear that the local maximum of the majority $p_z$ orbitals at $-0.75$~eV
which was already found to hybridize with the majority $d_{z^2}$ state is likewise affected by the rotation of magnetization at this energy. 
There are also changes in the $p_x$ states in this energy range as seen in the shaded area of Fig.~\ref{fig:FeAdatom}(c).
We will show below that the TAMR in the $p_z$ states can be explained by SOC induced mixing with the minority $p_x$
states. Close to $E_F$, the anisotropy of the LDOS is created by the 
majority $p_z$ and $d_{z^2}$ states as well whereas at higher energies above $+0.4$ eV only $p_z$ states exhibiting a negligible TAMR dominate the vacuum LDOS.
The TAMR value becomes large around +1~eV which is beyond the focus of the current investigation.

\subsubsection{TAMR of Mn adatom on Mn/W(110)}	
\begin{figure}[htb]
	\centering
	\includegraphics[scale=0.9,clip]{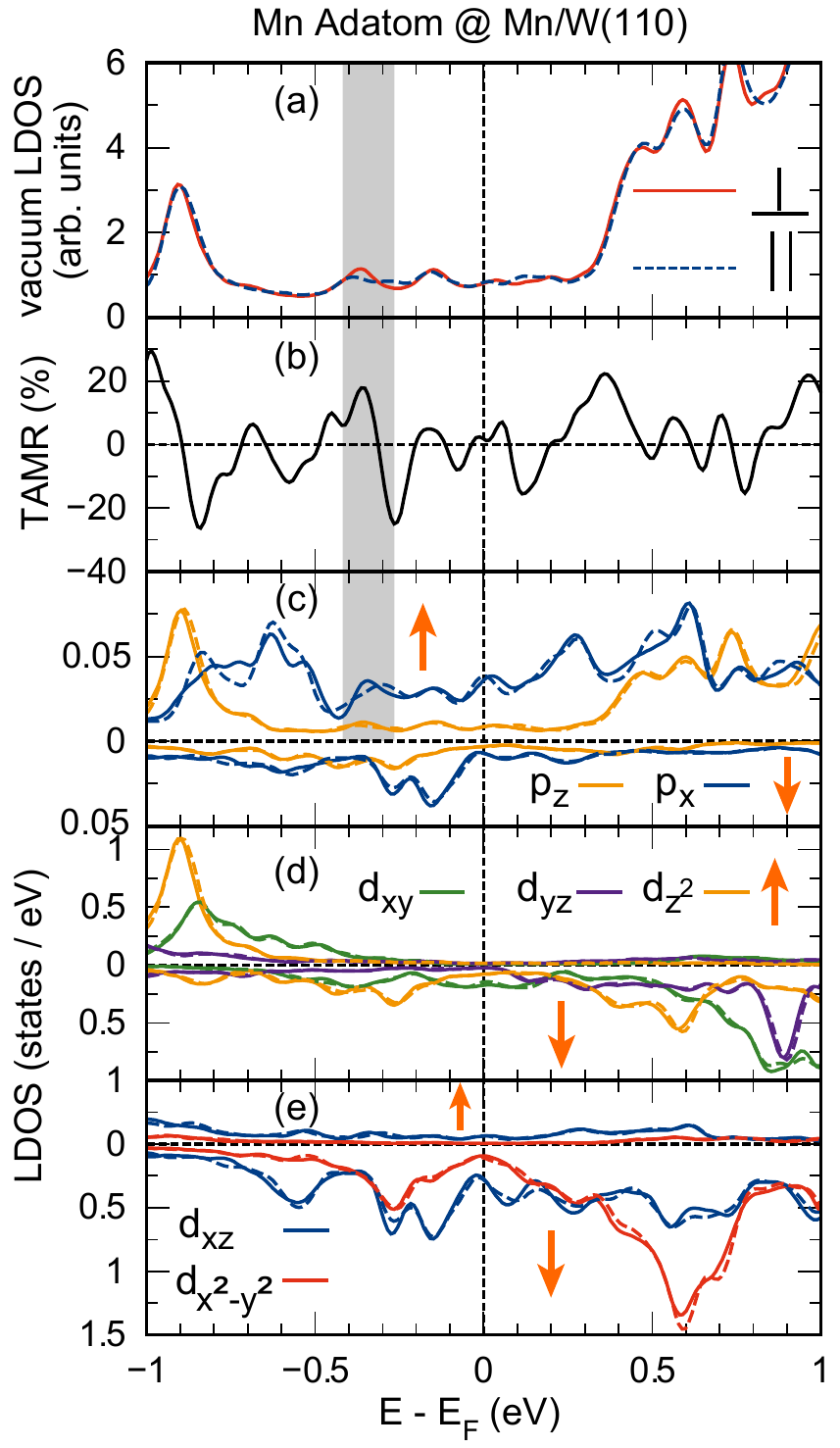}
	\caption{(a) LDOS in the vacuum calculated 6 {\AA} above the Mn adatom on Mn/W(110) for out-of-plane (solid red line) and in-plane (parallel to the $[1\overline{1}0]$ direction) magnetization directions (dashed blue line). (b) TAMR effect obtained from the data presented in (a) according to Eq.~(\ref{eq:TAMR}).(c), (d) and (e) orbital decomposition of the LDOS of the Mn adatom in terms of the majority (up) and minority (down) states as well as the orbital symmetry of the $p$- and $d$-states. Solid (dashed) lines correspond to the magnetization direction perpendicular (parallel) to the surface plane. The orange up and down arrows indicate majority- and minority-spin channels, respectively. Prominent peaks of the LDOS which will be investigated by means of a simple SOC model in section \ref{sec:tamr_model_result} are highlighted by shaded areas.}
	\label{fig:MnAdatom}
\end{figure}
Next we investigate the effect of SOC on the local density of states of a single magnetic Mn atom adsorbed on the Mn monolayer on W(110). In Fig.~\ref{fig:MnAdatom}(a) the spin-averaged vacuum LDOS calculated at 6 {\AA} above the Mn adatom is plotted for its two different magnetization directions. Again both curves almost coincide apart from a few barely noticeable differences occurring between $-0.9$ eV and $-0.8$ eV, $-0.35$ eV as well as $-0.26$ eV. These changes correspond to a moderate TAMR with maximum values ranging from $-26$\% up to +18\% below $E_F$ which is slightly higher than for the Fe adatom due to the larger spin splitting. 

A closer look at the orbital decomposition of the 
LDOS of the Mn adatom (see Figs.~\ref{fig:MnAdatom}(c), ~\ref{fig:MnAdatom}(d) and~\ref{fig:MnAdatom}(e)) reveals that the first peak in the TAMR at $-0.9$ eV originates from a shift of the dominant majority $d_{z^2}$ state towards higher energies upon rotation of the magnetization from out-of-plane to in-plane along the $[1\overline{1}0]$ direction. The same behavior can be observed for the majority $p_z$ state again at the same energetic position. The TAMR at the above mentioned other energies is due to interactions between $d_{z^2}$ and $d_{xz}$ states of both spin channels but also due to magnetization-direction dependent differences in the $p_z$ states 
[Fig.~\ref{fig:MnAdatom}(c)]. The anisotropy of the LDOS above $E_F$ taking values from $-18$\% to +20\% at 
+0.11~eV and +0.37~eV, respectively, can similarly be attributed to changes in the $d_{z^2}$ states of both spin directions. Above $+0.4$~eV contributions from the $p_z$ states come into play again 
as for the Fe adatom. 

We have also investigated the TAMR for the other $3d$ TM adatoms adsorbed on Mn/W(110) similar to the above mentioned cases (not shown). Our calculations reveal that the TAMR exhibits modest values for Ni ranging from $-24$\% up to +15\%. Although Ni might have an increased SOC strength compared to Fe and Co, the spin splitting between the orbitals that can interact via SOC is significantly smaller due to the small moment of 0.39 $\mu_B$. Analyzing the electronic structure influenced by SOC of a Cr and V adatom adsorbed on Mn/W(110) we found moderate values for the anisotropy of the LDOS as well. While the TAMR ranges from $-27$\% up to +16\% in the case of Cr, it shows values of $-17$\% up to +18\% for V. Thereby the V adatom exhibits the lowest anisotropy of its LDOS in comparison with all the other $3d$ elements studied in this paper which is in good agreement with its smallest SOC constant.  

\subsubsection{Modelling the TAMR of $3d$ adatoms}
\label{sec:tamr_model_result}
The TAMR of $3d$ transition-metal adatoms on Mn/W(110) can be explained using an analytically solvable tight-binding model~\cite{Neel2013}. This model is described by the localized surface atomic states interacting via SOC:
\begin{equation}
\label{eq:mod_tamr}
(E\cdot 1\!\!1 - H - \Sigma)G(E)=1\!\!1
\end{equation}
The diagonal elements of the Hamiltonian matrix, $H$, can be written as follows
\begin{equation}
H=\begin{pmatrix}
\epsilon_1 & -t \\
-t & \epsilon_2
\end{pmatrix}
\end{equation} 
where $\epsilon_1$ and $\epsilon_2$ are the energies of the two states and $t$ is the hopping term between them depending on the spin-quantization axis due to SOC. The non-Hermitian self-energy matrix, $\Sigma$, in Eq.~(\ref{eq:mod_tamr}) can be described as follows
\begin{equation}
\Sigma = \begin{pmatrix}
i\gamma_1 & 0 \\
0 & i\gamma_2
\end{pmatrix}
\end{equation}
The diagonal elements of the above matrix describe the broadening of the peaks arising due to the hybridization of the atomic states with the substrate. 
The diagonal elements of the Green's function matrix, $G_{11}$ and $G_{22}$, provide the LDOS, $n_1(E)$ and $n_2(E)$, of the two above mentioned states, respectively, and are given by 
\begin{equation}
n_i(E)=-\frac{1}{\pi}\text{Im}\left(G_{ii}(E)\right)
\end{equation}
\begin{table}[hbt]
\centering
\caption{Parameters (in eV) extracted from the DFT calculations 
for modeling the TAMR of an Fe and a Mn adatom on Mn/W(110) within the tight-binding model. 
$\gamma_1$ and  $\gamma_2$ describe the broadening of the two 
adatom states under study, whereas $\epsilon_1$ and $\epsilon_2$  are their respective energies. 
For the Fe
$d$ ($p$) states $\epsilon_1$ gives the energy of the majority $d_{z^2}$ ($p_z$) state and 
$\epsilon_2$ the minority $d_{xz}$ ($p_x$) state. For the Mn $p$ states $\epsilon_1$ denotes
the majority $p_x$ and $\epsilon_2$ the majority $p_z$ state.
$t_{\parallel}$ and $t_{\perp}$ represent the hopping term between them due to SOC for a magnetization direction
in-plane ($\parallel$) or out-of-plane ($\perp$).}
\label{tab:table3}
\begin{ruledtabular}
\begin{tabular}{c c c c c c}
 	& $\gamma_1$ & $\gamma_2$ & $\epsilon_1$ & $\epsilon_2$ & $t_{\parallel}$/$t_{\perp}$\\ 
	\colrule 
Fe $d$-states	& 0.066 &  0.070 & $-0.754$ &$-0.763$  & 0/0.035 \\ 
Fe $p$-states	&  0.070 & 0.090 & $-0.754$ & $-0.793$  & 0/0.040 \\ 
Mn $p$-states	& 0.060 & 0.107 & $-0.343$  & $-0.362$  & 0.040/0\\
\end{tabular}
\end{ruledtabular}
\end{table}

\begin{figure*}[htbp]
	\centering
	\includegraphics[scale=0.9,clip]{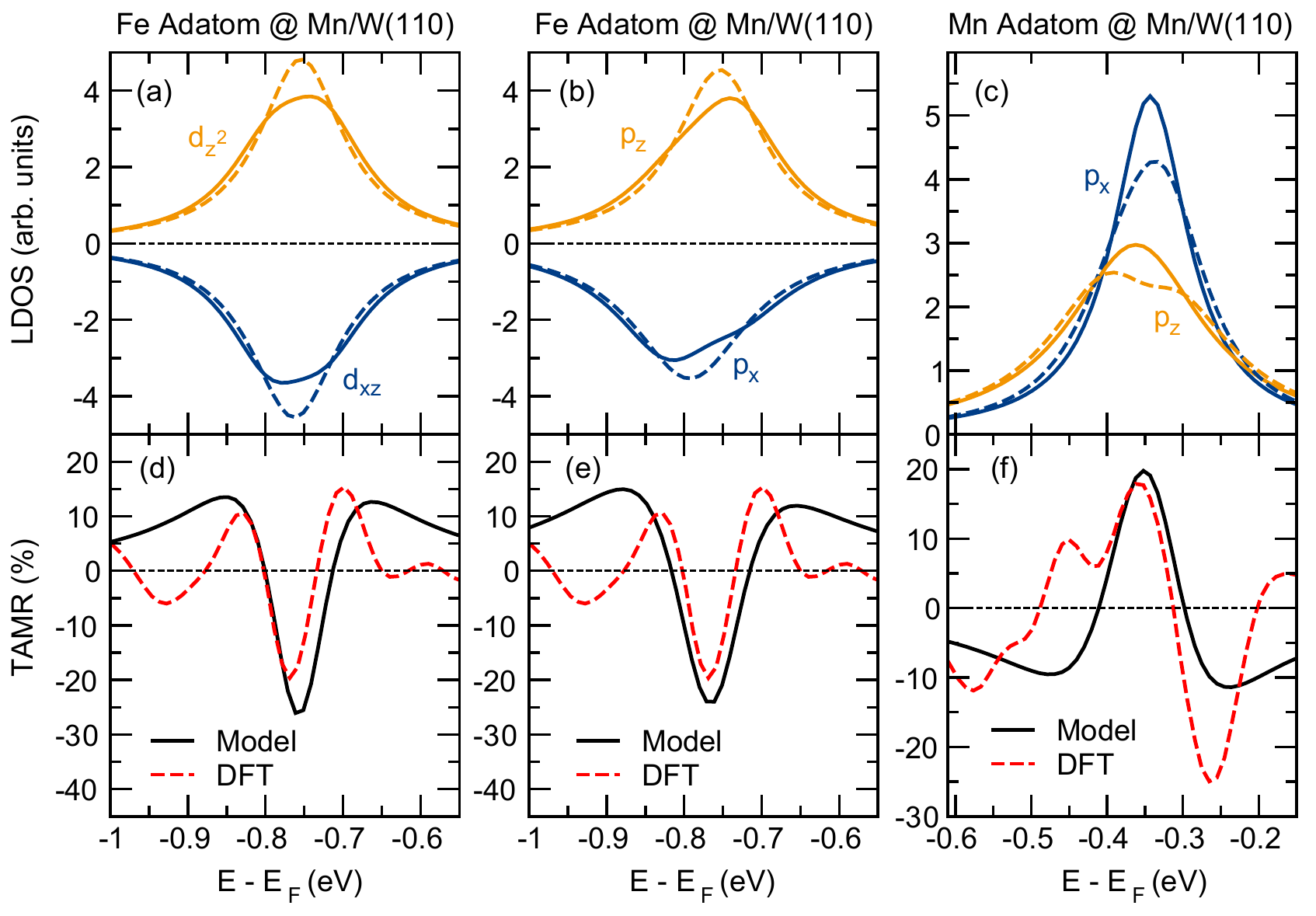}
	\caption{(a, b, c) LDOS calculated by means of the 
	simplified model of two localized atomic surface states with different orbital symmetry, $d_{z^2}$ 
	or $p_z$ (orange) and $d_{xz}$ or $p_x$ (blue), interacting via SOC for the case of (a, b) a Fe adatom and (c) a Mn adatom on Mn/W(110). The solid (dashed) lines correspond to the magnetization perpendicular (parallel along $[1\overline{1}0]$) to the surface plane. (d, e, f) Comparison of the TAMR calculated according to Eq.~(\ref{eq:TAMR}) for the $d_{z^2}$ state shown in (a) and the $p_z$ state shown in (b) and (c) (black solid lines) and the corresponding TAMR in the vacuum from the DFT calculations (red dashed lines).  }
	\label{TAMR_Model}
\end{figure*}
We use the Fe and the Mn adatom on Mn/W(110) as an example to illustrate the above mentioned model. Here, we consider two prominent peaks of states with different orbital symmetry and spin orientation from the LDOS of the DFT calculation, respectively. If both states are located nearly at the same energy 
and their curves show similar magnetization direction dependent changes, they are likely to hybridize by means of SOC
if the $H_{\text{SOC}}$ matrix element allows.

For the case of the Fe adatom, we examine the changes induced in the LDOS of the dominant majority $d_{z^2}$ and minority $d_{xz}$ state at approximately $-0.75$ eV upon rotation of the magnetization direction (see Figs.~\ref{fig:FeAdatom}(d) and~\ref{fig:FeAdatom}(e)). Their respective broadening and energetic positions are extracted from the data of the DFT calculation and listed in Table~\ref{tab:table3}. The dependence of the hopping term $t$ on the magnetization direction given by spherical coordinates ($\theta, \phi$), i.e. the mixing of states with the above mentioned symmetry, is given by the SOC matrix element~\cite{Abate1965}:
\begin{equation}
\Bigl|\langle \uparrow,d_z^2|H_{\text{SOC}}|d_{xz}, \downarrow\rangle \Bigr| = \frac{1}{2}\sqrt{3\left(\cos^2\phi+ \cos^2 \theta \sin^2 \phi\right)}
\label{eq:Matrixelement_Fe}
\end{equation}
with $H_{SOC}=\xi\textbf{L}\cdot \textbf{S}$ denoting the Hamiltonian of SOC, $\xi$ is the strength of this interaction.
$\textbf{L}$ and $\textbf{S}$ are the operators of the angular momentum and the spin, respectively. It vanishes for an in-plane magnetization along the $[1\overline{1}0]$ direction ($\phi$=90$^{\circ}$, $\theta$=90$^{\circ}$) and becomes maximal for the out-of-plane component ($\phi$=0$^{\circ}$, $\theta$=0$^{\circ}$). The strength of SOC for the $3d$ transition metals studied in this paper is in the order of 20 to 70 meV~\cite{StoehrSiegmann}, which is why the hopping term $t$ is taken as $t_\parallel = 0$ meV for the in-plane and $t_\perp = 35$ meV for the out-of-plane magnetization direction.

The result of this simplified model for the Fe adatom on Mn/W(110) is shown in Fig.~\ref{TAMR_Model}(a). It is evident that the model reproduces the LDOS of the two involved states calculated by means of DFT [cf. shaded areas in Figs.~\ref{fig:FeAdatom}(d) and~\ref{fig:FeAdatom}(e)] very well. The peaks of the modelled $d_{z^2}$ and $d_{xz}$ states are enhanced for the in-plane magnetization direction compared to the out-of-plane magnetization direction in agreement with the symmetry of the mixing parameter $t$ leading to a negative TAMR of $-26$\%.

In this model the anisotropy of the LDOS is evaluated by taking only the majority $d_{z^2}$ state into account due to its greater extent in the vacuum compared to the $d_{xz}$ state. Nonetheless the sign of the TAMR matches the results of DFT from the vacuum in an energy window between $-0.82$ eV and $-0.65$ eV which is highlighted by the red dashed line. Outside this small energy range the LDOS curve from the DFT calculation is much more complicated since it covers interactions from more than two states which cannot be described accordingly by this simplified model. As the model TAMR also neglects the influence of the minority $d_{z^2}$ states (as well as the $p_z$ orbitals) at this position, its magnitude exceeds the maximum value of $-20$\% from the DFT calculation which is lower due to canceling effects between the two spin channels. At this point we emphasize that the present system thereby shows not only a spin-orbit induced mixing of states with different orbital symmetry but additionally a spin mixing between majority and minority channel.

A similar model calculation can be applied to the majority $p_z$ state of the Fe adatom exhibiting the same magnetization-direction dependent changes of the LDOS at the same energy position (see shaded area in Fig.~\ref{fig:FeAdatom}(c)). The SOC matrix element describing an orbital and spin mixing between $p$ states with different symmetry takes 
the 
form~\cite{Haldar2019}:
\begin{equation}
\Bigl|\langle \uparrow,p_z|H_{\text{SOC}}|p_x, \downarrow\rangle \Bigr|  = 
\sqrt{\cos^2\phi + \cos^2 \theta \sin^2 \phi }{.}
\label{Matrixelement_Fe_p}
\end{equation}
As in the case of the matrix element for the $3d$ states, it vanishes for an in-plane magnetization direction and yields a non-zero value for the out-of-plane magnetization along the surface normal. 
Note that a mixing between $p_y$ and $p_z$ orbitals is not possible on the Mn/W(110) surface for the two considered magnetization directions under study due to the form of the matrix elements. The model calculation for the Fe $p$ states shown in Fig.~\ref{TAMR_Model}(b), 
using values of $t_\parallel = 0$~meV and $t_\perp = 40$~meV,
is again in good agreement with the DFT calculation from Fig.~\ref{fig:FeAdatom}(c). The peaks for both majority $p_z$ and minority $p_x$ states are enhanced for the in-plane magnetization compared to the out-of-plane component causing a negative TAMR that matches with the DFT result not only in sign and order of magnitude but also in energetic position (see Fig.~\ref{TAMR_Model}(e)). This observation supports the idea of a hybrid $p_z$-$d_z^2$ state. Under the influence of SOC it then also hybridizes with $p$ and $d$ states of different symmetry and as a consequence both types of orbitals contribute to the TAMR.

Similar to the case of a Fe adatom and consistent with previous observations in Section ~\ref{subsubsec:Mn_Adatom} concerning the spin polarization, the TAMR of a Mn adatom on the Mn/W(110) surface can be explained by magnetization direction dependent mixing of $p$ orbitals of the adatom dominating at energetic positions closer to $E_F$. Proceeding in the same way as before, we consider two peaks of majority $p_x$ and $p_z$ states at approximately $-0.35$ eV this time (see shaded areas in Figs.~\ref{fig:MnAdatom}(c)). Their broadening and energies calculated from the DFT are listed in Table~\ref{tab:table3} and the matrix element describing only orbital mixing via spin-orbit interaction between states of the mentioned orbitals is now 
given by \cite{Haldar2019}:
\begin{equation}
\Bigl|\langle \uparrow,p_z|H_{\text{SOC}}|p_x, \uparrow\rangle \Bigr|  = \sin\theta \sin \phi\text{.}
\label{Matrixelement_Mn}
\end{equation}

It vanishes for the out-of-plane magnetization direction ($\phi$=0$^{\circ}$, $\theta$=0$^{\circ}$) and becomes maximal for the in-plane magnetization direction aligned to the $[1\overline{1}0]$ direction ($\phi$=90$^{\circ}$, $\theta$=90$^{\circ}$). The strength of SOC is in the same order as for the Fe adatom. Therefore, the mixing parameter, $t$, is taken as $t_\perp = 0$ meV for the out-of-plane magnetization direction and $t_\parallel=40$ meV for the in-plane magnetization direction, respectively. The resulting model is illustrated in Fig.~\ref{TAMR_Model}(c). Again the LDOS of the two states is in good agreement with the results of the DFT calculation [cf. Figs.~\ref{fig:MnAdatom}(c)] with the peaks for the out-of-plane magnetization direction being enhanced and slightly shifted in energy compared to the in-plane magnetization direction in accordance with the values of the mixing parameter $t$. This behavior leads to a positive TAMR of +20\% at $-0.35$ eV for the $p_z$ state thereby 
slightly
exceeding the maximum value of +18\% at $-0.36$ eV calculated by DFT.

Consequently the simplified model of two localized atomic surface states demonstrates that the TAMR for Fe and Mn adatoms on Mn/W(110) is generated by a spin-orbit induced hybridization between $3p$ and $3d$ states of different orbital character and different spin for the case of the Fe adatom and $3p$ states with different orbital, but same spin character for the case of a Mn adatom.     

\section{Conclusion}
\label{sec:conclusion}
In summary, we have presented a systematic investigation of the electronic structure, magnetic properties, and spin polarization at the vacuum for the adsorption of single $3d$ TM adatoms on a Mn monolayer on W(110) by means of first-principles density functional theory calculations. Apart from that, we also studied the TAMR for the $3d$ TM adatoms adsorbed on Mn/W(110). The natural noncollinear magnetic ground state of the Mn/W(110) surface which takes the form of a cycloidal $173^{\circ}$ spin spiral has been approximated locally as a collinear antiferromagnet in order to examine its interaction with the adsorbed adatoms. The calculation of the TAMR has been restricted to two limiting cases of the magnetization direction of the respective adatom: (i) perpendicular to the surface (out-of-plane) and (i) parallel to the $[1\overline{1}0]$ (in-plane) direction which presents the propagation direction of the spin spiral. To benchmark our results, we have considered the adsorption of Co adatom on Mn/W(110). Our calculated results match well with the previously published results of Co adatom on Mn/W(110)~\cite{Caffrey2014a}.

Our analysis of the calculated electronic structure and magnetic properties indicates that the ferromagnetic coupling between the adsorbed $3d$ TM adatoms and the NN Mn atoms is energetically favorable in case of Mn, Fe, Co and Ni, whereas for V and Cr the antiferromagnetic alignment has lower energy. Additionally it becomes clear that the magnetic moments of the $3d$ TM adatoms follow the trend of Hund's first rule with Mn in the middle of the series exhibiting the largest moment. 
We also observe strong hybridization between the $3d$ TM adatoms and the Mn atoms of Mn/W(110) surface. Large values of the spin polarization of the LDOS at the surface as well as in the vacuum have also been observed due to the strong hybridization. While the $3d$ states of the adatom mainly contribute to this quantity further away from $E_F$ owing to their large spin splitting, $p_z$ and $s$ states dominate at $E_F$ in the vacuum causing high positive values of spin polarization up to 63\% for Mn, Fe, Co and Ni. Due to the AFM coupling with the NN Mn atoms, states of the same symmetry lead to large negative values in the spin polarization of up to $-73$\% in the case of Cr and V. Furthermore, differences with respect to the orbital occupation and spin splitting are clearly discernible by means of prominent peaks being shifted in the LDOS of the respective adatom.

Due to their small 
atomic number
all adatoms exhibit only a slight anisotropy of the LDOS leading to only modest values of the TAMR. In case of the Fe adatom, TAMR  rises up to 20\% at 0.75 eV below $E_F$ whereas for V adatom even smaller maximum values of 18\% were detected owing to the smallest 
atomic number
of all studied adatoms. The largest TAMR value of 27\% was obtained for the Cr adatom in accordance with its large magnetic moment.
In all cases the largest contribution to the TAMR can be attributed to a magnetization-direction dependent modification of $d_{z^2}$ and $p_z$ orbitals since states of this symmetry mainly preponderate in the vacuum.

The origin of the TAMR effect occurring for the $3d$ elements can be explained on the basis of a simplified model considering two localized atomic surface states of different orbital symmetry that interact via SOC. Especially for an Fe and a Mn adatom adsorbed on the Mn/W(110) surface it can be ascribed to a spin-orbit induced hybridization between 
majority $d_{z^2}$ and minority $d_{xz}$ states 
and majority $p_z$ and minority $p_x$ states
in the former case and to a hybridization mediated by SOC only between majority 
$p_x$ and $p_z$ states for the latter. 
In contrast to previous work on the TAMR of $3d$ transition-metal adatoms \cite{Neel2013,Caffrey2014a} we find both SOC 
induced mixing of opposite spin channels and a significant effect of $p$ states to the TAMR.

Furthermore, our analysis of the $3d$ transition-metal adatom partial charge densities in the vicinity of the Fermi energy have shown that the spin orientation of the respective adatom can be distinguished in SP-STM experiments at low bias voltages for the case of the V and Co adatom only.    

\section*{Acknowledgments} 
S.J.H. and S.H. acknowledge the Deutsche Forschungsgemeinschaft (DFG) via SFB677 for financial support. We gratefully acknowledge the computing time at the supercomputer of the North-German Supercomputing Alliance (HLRN). We thank N. M. Caffrey for valuable discussions. 

\input{reference.tex}

\end{document}

%% file: reference.tex
%

%% file: manuscript.bbl
\begin{thebibliography}{58}%
\makeatletter
\providecommand \@ifxundefined [1]{%
 \@ifx{#1\undefined}
}%
\providecommand \@ifnum [1]{%
 \ifnum #1\expandafter \@firstoftwo
 \else \expandafter \@secondoftwo
 \fi
}%
\providecommand \@ifx [1]{%
 \ifx #1\expandafter \@firstoftwo
 \else \expandafter \@secondoftwo
 \fi
}%
\providecommand \natexlab [1]{#1}%
\providecommand \enquote  [1]{``#1''}%
\providecommand \bibnamefont  [1]{#1}%
\providecommand \bibfnamefont [1]{#1}%
\providecommand \citenamefont [1]{#1}%
\providecommand \href@noop [0]{\@secondoftwo}%
\providecommand \href [0]{\begingroup \@sanitize@url \@href}%
\providecommand \@href[1]{\@@startlink{#1}\@@href}%
\providecommand \@@href[1]{\endgroup#1\@@endlink}%
\providecommand \@sanitize@url [0]{\catcode `\\12\catcode `\$12\catcode
  `\&12\catcode `\#12\catcode `\^12\catcode `\_12\catcode `\%12\relax}%
\providecommand \@@startlink[1]{}%
\providecommand \@@endlink[0]{}%
\providecommand \url  [0]{\begingroup\@sanitize@url \@url }%
\providecommand \@url [1]{\endgroup\@href {#1}{\urlprefix }}%
\providecommand \urlprefix  [0]{URL }%
\providecommand \Eprint [0]{\href }%
\providecommand \doibase [0]{https://doi.org/}%
\providecommand \selectlanguage [0]{\@gobble}%
\providecommand \bibinfo  [0]{\@secondoftwo}%
\providecommand \bibfield  [0]{\@secondoftwo}%
\providecommand \translation [1]{[#1]}%
\providecommand \BibitemOpen [0]{}%
\providecommand \bibitemStop [0]{}%
\providecommand \bibitemNoStop [0]{.\EOS\space}%
\providecommand \EOS [0]{\spacefactor3000\relax}%
\providecommand \BibitemShut  [1]{\csname bibitem#1\endcsname}%
\let\auto@bib@innerbib\@empty
\bibitem [{\citenamefont {Heide}\ \emph {et~al.}(2008)\citenamefont {Heide},
  \citenamefont {Bihlmayer},\ and\ \citenamefont {Bl{\"{u}}gel}}]{Heide2008}%
  \BibitemOpen
  \bibfield  {author} {\bibinfo {author} {\bibfnamefont {M.}~\bibnamefont
  {Heide}}, \bibinfo {author} {\bibfnamefont {G.}~\bibnamefont {Bihlmayer}},\
  and\ \bibinfo {author} {\bibfnamefont {S.}~\bibnamefont {Bl{\"{u}}gel}},\
  }\bibfield  {title} {\bibinfo {title} {{Dzyaloshinskii-Moriya interaction
  accounting for the orientation of magnetic domains in ultrathin films:
  Fe/W(110)}},\ }\href {https://doi.org/10.1103/PhysRevB.78.140403} {\bibfield
  {journal} {\bibinfo  {journal} {Phys. Rev. B}\ }\textbf {\bibinfo {volume}
  {78}},\ \bibinfo {pages} {140403} (\bibinfo {year} {2008})}\BibitemShut
  {NoStop}%
\bibitem [{\citenamefont {Ryu}\ \emph {et~al.}(2013)\citenamefont {Ryu},
  \citenamefont {Thomas}, \citenamefont {Yang},\ and\ \citenamefont
  {Parkin}}]{Ryu2013}%
  \BibitemOpen
  \bibfield  {author} {\bibinfo {author} {\bibfnamefont {K.-S.}\ \bibnamefont
  {Ryu}}, \bibinfo {author} {\bibfnamefont {L.}~\bibnamefont {Thomas}},
  \bibinfo {author} {\bibfnamefont {S.-H.}\ \bibnamefont {Yang}},\ and\
  \bibinfo {author} {\bibfnamefont {S.}~\bibnamefont {Parkin}},\ }\bibfield
  {title} {\bibinfo {title} {Chiral spin torque at magnetic domain walls},\
  }\href@noop {} {\bibfield  {journal} {\bibinfo  {journal} {Nature Nanotech.}\
  }\textbf {\bibinfo {volume} {8}},\ \bibinfo {pages} {527} (\bibinfo {year}
  {2013})}\BibitemShut {NoStop}%
\bibitem [{\citenamefont {Emori}\ \emph {et~al.}(2013)\citenamefont {Emori},
  \citenamefont {Bauer}, \citenamefont {Ahn}, \citenamefont {Martinez},\ and\
  \citenamefont {Beach}}]{Emori2013}%
  \BibitemOpen
  \bibfield  {author} {\bibinfo {author} {\bibfnamefont {S.}~\bibnamefont
  {Emori}}, \bibinfo {author} {\bibfnamefont {U.}~\bibnamefont {Bauer}},
  \bibinfo {author} {\bibfnamefont {S.-M.}\ \bibnamefont {Ahn}}, \bibinfo
  {author} {\bibfnamefont {E.}~\bibnamefont {Martinez}},\ and\ \bibinfo
  {author} {\bibfnamefont {G.~S.~D.}\ \bibnamefont {Beach}},\ }\bibfield
  {title} {\bibinfo {title} {Current-driven dynamics of chiral ferromagnetic
  domain walls},\ }\href@noop {} {\bibfield  {journal} {\bibinfo  {journal}
  {Nature Mat.}\ }\textbf {\bibinfo {volume} {12}},\ \bibinfo {pages} {611}
  (\bibinfo {year} {2013})}\BibitemShut {NoStop}%
\bibitem [{\citenamefont {Bogdanov}\ and\ \citenamefont
  {Yablonskii}(1989)}]{Bogdanov1989}%
  \BibitemOpen
  \bibfield  {author} {\bibinfo {author} {\bibfnamefont {A.}~\bibnamefont
  {Bogdanov}}\ and\ \bibinfo {author} {\bibfnamefont {D.~A.}\ \bibnamefont
  {Yablonskii}},\ }\bibfield  {title} {\bibinfo {title} {Thermodynamically
  stable ‘vortices’ in magnetically ordered crystals. the mixed state of
  magnets.},\ }\href@noop {} {\bibfield  {journal} {\bibinfo  {journal} {Sov.
  Phys. JETP}\ }\textbf {\bibinfo {volume} {68}},\ \bibinfo {pages} {101}
  (\bibinfo {year} {1989})}\BibitemShut {NoStop}%
\bibitem [{\citenamefont {Bogdanov}\ and\ \citenamefont
  {R{\"o}{\ss}ler}(2001)}]{Bogdanov:2001aa}%
  \BibitemOpen
  \bibfield  {author} {\bibinfo {author} {\bibfnamefont {A.~N.}\ \bibnamefont
  {Bogdanov}}\ and\ \bibinfo {author} {\bibfnamefont {U.~K.}\ \bibnamefont
  {R{\"o}{\ss}ler}},\ }\bibfield  {title} {\bibinfo {title} {Chiral symmetry
  breaking in magnetic thin films and multilayers},\ }\href@noop {} {\bibfield
  {journal} {\bibinfo  {journal} {Phys. Rev. Lett.}\ }\textbf {\bibinfo
  {volume} {87}},\ \bibinfo {pages} {037203} (\bibinfo {year}
  {2001})}\BibitemShut {NoStop}%
\bibitem [{\citenamefont {Heinze}\ \emph {et~al.}(2011)\citenamefont {Heinze},
  \citenamefont {von Bergmann}, \citenamefont {Menzel}, \citenamefont {Brede},
  \citenamefont {Kubetzka}, \citenamefont {Wiesendanger}, \citenamefont
  {Bihlmayer},\ and\ \citenamefont {Bl{\"u}gel}}]{Heinze2011}%
  \BibitemOpen
  \bibfield  {author} {\bibinfo {author} {\bibfnamefont {S.}~\bibnamefont
  {Heinze}}, \bibinfo {author} {\bibfnamefont {K.}~\bibnamefont {von
  Bergmann}}, \bibinfo {author} {\bibfnamefont {M.}~\bibnamefont {Menzel}},
  \bibinfo {author} {\bibfnamefont {J.}~\bibnamefont {Brede}}, \bibinfo
  {author} {\bibfnamefont {A.}~\bibnamefont {Kubetzka}}, \bibinfo {author}
  {\bibfnamefont {R.}~\bibnamefont {Wiesendanger}}, \bibinfo {author}
  {\bibfnamefont {G.}~\bibnamefont {Bihlmayer}},\ and\ \bibinfo {author}
  {\bibfnamefont {S.}~\bibnamefont {Bl{\"u}gel}},\ }\bibfield  {title}
  {\bibinfo {title} {Spontaneous atomic-scale magnetic skyrmion lattice in two
  dimensions},\ }\href@noop {} {\bibfield  {journal} {\bibinfo  {journal} {Nat.
  Phys.}\ }\textbf {\bibinfo {volume} {7}},\ \bibinfo {pages} {713} (\bibinfo
  {year} {2011})}\BibitemShut {NoStop}%
\bibitem [{\citenamefont {Romming}\ \emph {et~al.}(2013)\citenamefont
  {Romming}, \citenamefont {Hanneken}, \citenamefont {Menzel}, \citenamefont
  {Bickel}, \citenamefont {Wolter}, \citenamefont {von Bergmann}, \citenamefont
  {Kubetzka},\ and\ \citenamefont {Wiesendanger}}]{Romming2013}%
  \BibitemOpen
  \bibfield  {author} {\bibinfo {author} {\bibfnamefont {N.}~\bibnamefont
  {Romming}}, \bibinfo {author} {\bibfnamefont {C.}~\bibnamefont {Hanneken}},
  \bibinfo {author} {\bibfnamefont {M.}~\bibnamefont {Menzel}}, \bibinfo
  {author} {\bibfnamefont {J.~E.}\ \bibnamefont {Bickel}}, \bibinfo {author}
  {\bibfnamefont {B.}~\bibnamefont {Wolter}}, \bibinfo {author} {\bibfnamefont
  {K.}~\bibnamefont {von Bergmann}}, \bibinfo {author} {\bibfnamefont
  {A.}~\bibnamefont {Kubetzka}},\ and\ \bibinfo {author} {\bibfnamefont
  {R.}~\bibnamefont {Wiesendanger}},\ }\bibfield  {title} {\bibinfo {title}
  {Writing and deleting single magnetic skyrmions},\ }\href@noop {} {\bibfield
  {journal} {\bibinfo  {journal} {Science}\ }\textbf {\bibinfo {volume}
  {341}},\ \bibinfo {pages} {636} (\bibinfo {year} {2013})}\BibitemShut
  {NoStop}%
\bibitem [{\citenamefont {Herv{\'{e}}}\ \emph {et~al.}(2018)\citenamefont
  {Herv{\'{e}}}, \citenamefont {Dup{\'{e}}}, \citenamefont {Lopes},
  \citenamefont {B{\"{o}}ttcher}, \citenamefont {Martins}, \citenamefont
  {Balashov}, \citenamefont {Gerhard}, \citenamefont {Sinova},\ and\
  \citenamefont {Wulfhekel}}]{Herve2018}%
  \BibitemOpen
  \bibfield  {author} {\bibinfo {author} {\bibfnamefont {M.}~\bibnamefont
  {Herv{\'{e}}}}, \bibinfo {author} {\bibfnamefont {B.}~\bibnamefont
  {Dup{\'{e}}}}, \bibinfo {author} {\bibfnamefont {R.}~\bibnamefont {Lopes}},
  \bibinfo {author} {\bibfnamefont {M.}~\bibnamefont {B{\"{o}}ttcher}},
  \bibinfo {author} {\bibfnamefont {M.~D.}\ \bibnamefont {Martins}}, \bibinfo
  {author} {\bibfnamefont {T.}~\bibnamefont {Balashov}}, \bibinfo {author}
  {\bibfnamefont {L.}~\bibnamefont {Gerhard}}, \bibinfo {author} {\bibfnamefont
  {J.}~\bibnamefont {Sinova}},\ and\ \bibinfo {author} {\bibfnamefont
  {W.}~\bibnamefont {Wulfhekel}},\ }\bibfield  {title} {\bibinfo {title}
  {{Stabilizing spin spirals and isolated skyrmions at low magnetic field
  exploiting vanishing magnetic anisotropy}},\ }\href
  {https://doi.org/10.1038/s41467-018-03240-w} {\bibfield  {journal} {\bibinfo
  {journal} {Nature Communications}\ }\textbf {\bibinfo {volume} {9}},\
  \bibinfo {pages} {1015} (\bibinfo {year} {2018})}\BibitemShut {NoStop}%
\bibitem [{\citenamefont {Ferriani}\ \emph {et~al.}(2008)\citenamefont
  {Ferriani}, \citenamefont {von Bergmann}, \citenamefont {Vedmedenko},
  \citenamefont {Heinze}, \citenamefont {Bode}, \citenamefont {Heide},
  \citenamefont {Bihlmayer}, \citenamefont {Bl\"ugel},\ and\ \citenamefont
  {Wiesendanger}}]{Ferriani2008}%
  \BibitemOpen
  \bibfield  {author} {\bibinfo {author} {\bibfnamefont {P.}~\bibnamefont
  {Ferriani}}, \bibinfo {author} {\bibfnamefont {K.}~\bibnamefont {von
  Bergmann}}, \bibinfo {author} {\bibfnamefont {E.~Y.}\ \bibnamefont
  {Vedmedenko}}, \bibinfo {author} {\bibfnamefont {S.}~\bibnamefont {Heinze}},
  \bibinfo {author} {\bibfnamefont {M.}~\bibnamefont {Bode}}, \bibinfo {author}
  {\bibfnamefont {M.}~\bibnamefont {Heide}}, \bibinfo {author} {\bibfnamefont
  {G.}~\bibnamefont {Bihlmayer}}, \bibinfo {author} {\bibfnamefont
  {S.}~\bibnamefont {Bl\"ugel}},\ and\ \bibinfo {author} {\bibfnamefont
  {R.}~\bibnamefont {Wiesendanger}},\ }\bibfield  {title} {\bibinfo {title}
  {Atomic-scale spin spiral with a unique rotational sense: {Mn} monolayer on
  {W(001)}},\ }\href@noop {} {\bibfield  {journal} {\bibinfo  {journal} {Phys.
  Rev. Lett.}\ }\textbf {\bibinfo {volume} {101}},\ \bibinfo {pages} {027201}
  (\bibinfo {year} {2008})}\BibitemShut {NoStop}%
\bibitem [{\citenamefont {Phark}\ \emph {et~al.}(2014)\citenamefont {Phark},
  \citenamefont {Fischer}, \citenamefont {Corbetta}, \citenamefont {Sander},
  \citenamefont {Nakamura},\ and\ \citenamefont {Kirschner}}]{Phark2014}%
  \BibitemOpen
  \bibfield  {author} {\bibinfo {author} {\bibfnamefont {S.-H.}\ \bibnamefont
  {Phark}}, \bibinfo {author} {\bibfnamefont {J.~A.}\ \bibnamefont {Fischer}},
  \bibinfo {author} {\bibfnamefont {M.}~\bibnamefont {Corbetta}}, \bibinfo
  {author} {\bibfnamefont {D.}~\bibnamefont {Sander}}, \bibinfo {author}
  {\bibfnamefont {K.}~\bibnamefont {Nakamura}},\ and\ \bibinfo {author}
  {\bibfnamefont {J.}~\bibnamefont {Kirschner}},\ }\bibfield  {title} {\bibinfo
  {title} {Reduced-dimensionality-induced helimagnetism in iron nanoislands},\
  }\href@noop {} {\bibfield  {journal} {\bibinfo  {journal} {Nature Comm.}\
  }\textbf {\bibinfo {volume} {5}},\ \bibinfo {pages} {5183} (\bibinfo {year}
  {2014})}\BibitemShut {NoStop}%
\bibitem [{\citenamefont {Fert}\ \emph {et~al.}(2013)\citenamefont {Fert},
  \citenamefont {Cros},\ and\ \citenamefont {Sampaio}}]{Fert2013}%
  \BibitemOpen
  \bibfield  {author} {\bibinfo {author} {\bibfnamefont {A.}~\bibnamefont
  {Fert}}, \bibinfo {author} {\bibfnamefont {V.}~\bibnamefont {Cros}},\ and\
  \bibinfo {author} {\bibfnamefont {J.}~\bibnamefont {Sampaio}},\ }\bibfield
  {title} {\bibinfo {title} {{Skyrmions on the track}},\ }\href@noop {}
  {\bibfield  {journal} {\bibinfo  {journal} {Nat. Nano.}\ }\textbf {\bibinfo
  {volume} {8}},\ \bibinfo {pages} {152} (\bibinfo {year} {2013})}\BibitemShut
  {NoStop}%
\bibitem [{\citenamefont {Nagaosa}\ and\ \citenamefont
  {Tokura}(2013)}]{Nagaosa2013}%
  \BibitemOpen
  \bibfield  {author} {\bibinfo {author} {\bibfnamefont {N.}~\bibnamefont
  {Nagaosa}}\ and\ \bibinfo {author} {\bibfnamefont {Y.}~\bibnamefont
  {Tokura}},\ }\bibfield  {title} {\bibinfo {title} {Topological properties and
  dynamics of magnetic skyrmions},\ }\href@noop {} {\bibfield  {journal}
  {\bibinfo  {journal} {Nat. Nano.}\ }\textbf {\bibinfo {volume} {8}},\
  \bibinfo {pages} {899} (\bibinfo {year} {2013})}\BibitemShut {NoStop}%
\bibitem [{\citenamefont {Bode}\ \emph {et~al.}(2007)\citenamefont {Bode},
  \citenamefont {Heide}, \citenamefont {von Bergmann}, \citenamefont
  {Ferriani}, \citenamefont {Heinze}, \citenamefont {Bihlmayer}, \citenamefont
  {Kubetzka}, \citenamefont {Pietzsch}, \citenamefont {Bl{\"u}gel},\ and\
  \citenamefont {Wiesendanger}}]{Bode2007}%
  \BibitemOpen
  \bibfield  {author} {\bibinfo {author} {\bibfnamefont {M.}~\bibnamefont
  {Bode}}, \bibinfo {author} {\bibfnamefont {M.}~\bibnamefont {Heide}},
  \bibinfo {author} {\bibfnamefont {K.}~\bibnamefont {von Bergmann}}, \bibinfo
  {author} {\bibfnamefont {P.}~\bibnamefont {Ferriani}}, \bibinfo {author}
  {\bibfnamefont {S.}~\bibnamefont {Heinze}}, \bibinfo {author} {\bibfnamefont
  {G.}~\bibnamefont {Bihlmayer}}, \bibinfo {author} {\bibfnamefont
  {A.}~\bibnamefont {Kubetzka}}, \bibinfo {author} {\bibfnamefont
  {O.}~\bibnamefont {Pietzsch}}, \bibinfo {author} {\bibfnamefont
  {S.}~\bibnamefont {Bl{\"u}gel}},\ and\ \bibinfo {author} {\bibfnamefont
  {R.}~\bibnamefont {Wiesendanger}},\ }\bibfield  {title} {\bibinfo {title}
  {Chiral magnetic order at surfaces driven by inversion asymmetry},\
  }\href@noop {} {\bibfield  {journal} {\bibinfo  {journal} {Nature}\ }\textbf
  {\bibinfo {volume} {447}},\ \bibinfo {pages} {190} (\bibinfo {year}
  {2007})}\BibitemShut {NoStop}%
\bibitem [{\citenamefont {Lounis}\ \emph {et~al.}(2007)\citenamefont {Lounis},
  \citenamefont {Mavropoulos}, \citenamefont {Zeller}, \citenamefont
  {Dederichs},\ and\ \citenamefont {Bl\"ugel}}]{Lounis2007}%
  \BibitemOpen
  \bibfield  {author} {\bibinfo {author} {\bibfnamefont {S.}~\bibnamefont
  {Lounis}}, \bibinfo {author} {\bibfnamefont {P.}~\bibnamefont {Mavropoulos}},
  \bibinfo {author} {\bibfnamefont {R.}~\bibnamefont {Zeller}}, \bibinfo
  {author} {\bibfnamefont {P.~H.}\ \bibnamefont {Dederichs}},\ and\ \bibinfo
  {author} {\bibfnamefont {S.}~\bibnamefont {Bl\"ugel}},\ }\bibfield  {title}
  {\bibinfo {title} {Noncollinear magnetism of {Cr} and {Mn} nanoclusters on
  {Ni(111)}: Changing the magnetic configuration atom by atom},\ }\href
  {https://doi.org/10.1103/PhysRevB.75.174436} {\bibfield  {journal} {\bibinfo
  {journal} {Phys. Rev. B}\ }\textbf {\bibinfo {volume} {75}},\ \bibinfo
  {pages} {174436} (\bibinfo {year} {2007})}\BibitemShut {NoStop}%
\bibitem [{\citenamefont {Enders}\ \emph {et~al.}(2010)\citenamefont {Enders},
  \citenamefont {Skomski},\ and\ \citenamefont {Honolka}}]{Enders2010}%
  \BibitemOpen
  \bibfield  {author} {\bibinfo {author} {\bibfnamefont {A.}~\bibnamefont
  {Enders}}, \bibinfo {author} {\bibfnamefont {R.}~\bibnamefont {Skomski}},\
  and\ \bibinfo {author} {\bibfnamefont {J.}~\bibnamefont {Honolka}},\
  }\bibfield  {title} {\bibinfo {title} {Magnetic surface nanostructures},\
  }\href {https://doi.org/10.1088/0953-8984/22/43/433001} {\bibfield  {journal}
  {\bibinfo  {journal} {Journal of Physics: Condensed Matter}\ }\textbf
  {\bibinfo {volume} {22}},\ \bibinfo {pages} {433001} (\bibinfo {year}
  {2010})}\BibitemShut {NoStop}%
\bibitem [{\citenamefont {Hermenau}\ \emph {et~al.}(2017)\citenamefont
  {Hermenau}, \citenamefont {Iba{\~n}ez-Azpiroz}, \citenamefont {H{\"u}bner},
  \citenamefont {Sonntag}, \citenamefont {Baxevanis}, \citenamefont {Ton},
  \citenamefont {Steinbrecher}, \citenamefont {Khajetoorians}, \citenamefont
  {dos Santos~Dias}, \citenamefont {Bl{\"u}gel}, \citenamefont {Wiesendanger},
  \citenamefont {Lounis},\ and\ \citenamefont {Wiebe}}]{Hermenau2017}%
  \BibitemOpen
  \bibfield  {author} {\bibinfo {author} {\bibfnamefont {J.}~\bibnamefont
  {Hermenau}}, \bibinfo {author} {\bibfnamefont {J.}~\bibnamefont
  {Iba{\~n}ez-Azpiroz}}, \bibinfo {author} {\bibfnamefont {C.}~\bibnamefont
  {H{\"u}bner}}, \bibinfo {author} {\bibfnamefont {A.}~\bibnamefont {Sonntag}},
  \bibinfo {author} {\bibfnamefont {B.}~\bibnamefont {Baxevanis}}, \bibinfo
  {author} {\bibfnamefont {K.~T.}\ \bibnamefont {Ton}}, \bibinfo {author}
  {\bibfnamefont {M.}~\bibnamefont {Steinbrecher}}, \bibinfo {author}
  {\bibfnamefont {A.~A.}\ \bibnamefont {Khajetoorians}}, \bibinfo {author}
  {\bibfnamefont {M.}~\bibnamefont {dos Santos~Dias}}, \bibinfo {author}
  {\bibfnamefont {S.}~\bibnamefont {Bl{\"u}gel}}, \bibinfo {author}
  {\bibfnamefont {R.}~\bibnamefont {Wiesendanger}}, \bibinfo {author}
  {\bibfnamefont {S.}~\bibnamefont {Lounis}},\ and\ \bibinfo {author}
  {\bibfnamefont {J.}~\bibnamefont {Wiebe}},\ }\bibfield  {title} {\bibinfo
  {title} {A gateway towards non-collinear spin processing using three-atom
  magnets with strong substrate coupling},\ }\href
  {https://doi.org/10.1038/s41467-017-00506-7} {\bibfield  {journal} {\bibinfo
  {journal} {Nature Communications}\ }\textbf {\bibinfo {volume} {8}},\
  \bibinfo {pages} {642} (\bibinfo {year} {2017})}\BibitemShut {NoStop}%
\bibitem [{\citenamefont {Wiesendanger}\ \emph {et~al.}(1990)\citenamefont
  {Wiesendanger}, \citenamefont {G\"untherodt}, \citenamefont {G\"untherodt},
  \citenamefont {Gambino},\ and\ \citenamefont {Ruf}}]{Wiesendanger1990}%
  \BibitemOpen
  \bibfield  {author} {\bibinfo {author} {\bibfnamefont {R.}~\bibnamefont
  {Wiesendanger}}, \bibinfo {author} {\bibfnamefont {H.-J.}\ \bibnamefont
  {G\"untherodt}}, \bibinfo {author} {\bibfnamefont {G.}~\bibnamefont
  {G\"untherodt}}, \bibinfo {author} {\bibfnamefont {R.~J.}\ \bibnamefont
  {Gambino}},\ and\ \bibinfo {author} {\bibfnamefont {R.}~\bibnamefont {Ruf}},\
  }\bibfield  {title} {\bibinfo {title} {Observation of vacuum tunneling of
  spin-polarized electrons with the scanning tunneling microscope},\ }\href
  {https://doi.org/10.1103/PhysRevLett.65.247} {\bibfield  {journal} {\bibinfo
  {journal} {Phys. Rev. Lett.}\ }\textbf {\bibinfo {volume} {65}},\ \bibinfo
  {pages} {247} (\bibinfo {year} {1990})}\BibitemShut {NoStop}%
\bibitem [{\citenamefont {Wulfhekel}\ and\ \citenamefont
  {Kirschner}(1999)}]{Wulfhekel1999}%
  \BibitemOpen
  \bibfield  {author} {\bibinfo {author} {\bibfnamefont {W.}~\bibnamefont
  {Wulfhekel}}\ and\ \bibinfo {author} {\bibfnamefont {J.}~\bibnamefont
  {Kirschner}},\ }\bibfield  {title} {\bibinfo {title} {Spin-polarized scanning
  tunneling microscopy on ferromagnets},\ }\href
  {https://doi.org/10.1063/1.124879} {\bibfield  {journal} {\bibinfo  {journal}
  {Applied Physics Letters}\ }\textbf {\bibinfo {volume} {75}},\ \bibinfo
  {pages} {1944} (\bibinfo {year} {1999})}\BibitemShut {NoStop}%
\bibitem [{\citenamefont {Bode}(2003)}]{Bode2003}%
  \BibitemOpen
  \bibfield  {author} {\bibinfo {author} {\bibfnamefont {M.}~\bibnamefont
  {Bode}},\ }\bibfield  {title} {\bibinfo {title} {Spin-polarized scanning
  tunneling microscopy},\ }\href@noop {} {\bibfield  {journal} {\bibinfo
  {journal} {Rep. Prog. Phys.}\ }\textbf {\bibinfo {volume} {66}},\ \bibinfo
  {pages} {523} (\bibinfo {year} {2003})}\BibitemShut {NoStop}%
\bibitem [{\citenamefont {Hauptmann}\ \emph {et~al.}(2017)\citenamefont
  {Hauptmann}, \citenamefont {Gerritsen}, \citenamefont {Wegner},\ and\
  \citenamefont {Khajetoorians}}]{Hauptmann2017}%
  \BibitemOpen
  \bibfield  {author} {\bibinfo {author} {\bibfnamefont {N.}~\bibnamefont
  {Hauptmann}}, \bibinfo {author} {\bibfnamefont {J.~W.}\ \bibnamefont
  {Gerritsen}}, \bibinfo {author} {\bibfnamefont {D.}~\bibnamefont {Wegner}},\
  and\ \bibinfo {author} {\bibfnamefont {A.~A.}\ \bibnamefont
  {Khajetoorians}},\ }\bibfield  {title} {\bibinfo {title} {Sensing
  noncollinear magnetism at the atomic scale combining magnetic exchange and
  spin-polarized imaging},\ }\href
  {https://doi.org/10.1021/acs.nanolett.7b02538} {\bibfield  {journal}
  {\bibinfo  {journal} {Nano Letters}\ }\textbf {\bibinfo {volume} {17}},\
  \bibinfo {pages} {5660} (\bibinfo {year} {2017})}\BibitemShut {NoStop}%
\bibitem [{\citenamefont {Hauptmann}\ \emph {et~al.}(2020)\citenamefont
  {Hauptmann}, \citenamefont {Haldar}, \citenamefont {Hung}, \citenamefont
  {Jolie}, \citenamefont {Gutzeit}, \citenamefont {Wegner}, \citenamefont
  {Heinze},\ and\ \citenamefont {Khajetoorians}}]{Hauptmann2019}%
  \BibitemOpen
  \bibfield  {author} {\bibinfo {author} {\bibfnamefont {N.}~\bibnamefont
  {Hauptmann}}, \bibinfo {author} {\bibfnamefont {S.}~\bibnamefont {Haldar}},
  \bibinfo {author} {\bibfnamefont {T.-C.}\ \bibnamefont {Hung}}, \bibinfo
  {author} {\bibfnamefont {W.}~\bibnamefont {Jolie}}, \bibinfo {author}
  {\bibfnamefont {M.}~\bibnamefont {Gutzeit}}, \bibinfo {author} {\bibfnamefont
  {D.}~\bibnamefont {Wegner}}, \bibinfo {author} {\bibfnamefont
  {S.}~\bibnamefont {Heinze}},\ and\ \bibinfo {author} {\bibfnamefont {A.~A.}\
  \bibnamefont {Khajetoorians}},\ }\bibfield  {title} {\bibinfo {title}
  {Quantifying exchange forces of a non-collinear magnetic structure on the
  atomic scale},\ }\href@noop {} {\bibfield  {journal} {\bibinfo  {journal}
  {Nature Communications}\ }\textbf {\bibinfo {volume} {11}},\ \bibinfo {pages}
  {1197} (\bibinfo {year} {2020})}\BibitemShut {NoStop}%
\bibitem [{\citenamefont {Serrate}\ \emph {et~al.}(2010)\citenamefont
  {Serrate}, \citenamefont {Ferriani}, \citenamefont {Yoshida}, \citenamefont
  {Hla}, \citenamefont {Menzel}, \citenamefont {von Bergmann}, \citenamefont
  {Heinze}, \citenamefont {Kubetzka},\ and\ \citenamefont
  {Wiesendanger}}]{Serrate2010}%
  \BibitemOpen
  \bibfield  {author} {\bibinfo {author} {\bibfnamefont {D.}~\bibnamefont
  {Serrate}}, \bibinfo {author} {\bibfnamefont {P.}~\bibnamefont {Ferriani}},
  \bibinfo {author} {\bibfnamefont {Y.}~\bibnamefont {Yoshida}}, \bibinfo
  {author} {\bibfnamefont {S.-W.}\ \bibnamefont {Hla}}, \bibinfo {author}
  {\bibfnamefont {M.}~\bibnamefont {Menzel}}, \bibinfo {author} {\bibfnamefont
  {K.}~\bibnamefont {von Bergmann}}, \bibinfo {author} {\bibfnamefont
  {S.}~\bibnamefont {Heinze}}, \bibinfo {author} {\bibfnamefont
  {A.}~\bibnamefont {Kubetzka}},\ and\ \bibinfo {author} {\bibfnamefont
  {R.}~\bibnamefont {Wiesendanger}},\ }\bibfield  {title} {\bibinfo {title}
  {Imaging and manipulating the spin direction of individual atoms},\
  }\href@noop {} {\bibfield  {journal} {\bibinfo  {journal} {Nature Nanotech.}\
  }\textbf {\bibinfo {volume} {5}},\ \bibinfo {pages} {350} (\bibinfo {year}
  {2010})}\BibitemShut {NoStop}%
\bibitem [{\citenamefont {Serrate}\ \emph {et~al.}(2016)\citenamefont
  {Serrate}, \citenamefont {Yoshida}, \citenamefont {Moro-Lagares},
  \citenamefont {Kubetzka},\ and\ \citenamefont {Wiesendanger}}]{Serrate2016}%
  \BibitemOpen
  \bibfield  {author} {\bibinfo {author} {\bibfnamefont {D.}~\bibnamefont
  {Serrate}}, \bibinfo {author} {\bibfnamefont {Y.}~\bibnamefont {Yoshida}},
  \bibinfo {author} {\bibfnamefont {M.}~\bibnamefont {Moro-Lagares}}, \bibinfo
  {author} {\bibfnamefont {A.}~\bibnamefont {Kubetzka}},\ and\ \bibinfo
  {author} {\bibfnamefont {R.}~\bibnamefont {Wiesendanger}},\ }\bibfield
  {title} {\bibinfo {title} {Spin-sensitive shape asymmetry of adatoms on
  noncollinear magnetic substrates},\ }\href@noop {} {\bibfield  {journal}
  {\bibinfo  {journal} {Phys. Rev. B}\ }\textbf {\bibinfo {volume} {93}},\
  \bibinfo {pages} {125424} (\bibinfo {year} {2016})}\BibitemShut {NoStop}%
\bibitem [{\citenamefont {Haldar}\ and\ \citenamefont
  {Heinze}(2018)}]{Haldar2018}%
  \BibitemOpen
  \bibfield  {author} {\bibinfo {author} {\bibfnamefont {S.}~\bibnamefont
  {Haldar}}\ and\ \bibinfo {author} {\bibfnamefont {S.}~\bibnamefont
  {Heinze}},\ }\bibfield  {title} {\bibinfo {title} {{Noncollinear spin density
  of an adatom on a magnetic surface}},\ }\href
  {https://doi.org/10.1103/PhysRevB.98.220401} {\bibfield  {journal} {\bibinfo
  {journal} {Phys. Rev. B}\ }\textbf {\bibinfo {volume} {98}},\ \bibinfo
  {pages} {220401(R)} (\bibinfo {year} {2018})}\BibitemShut {NoStop}%
\bibitem [{\citenamefont {Ferriani}\ \emph {et~al.}(2010)\citenamefont
  {Ferriani}, \citenamefont {Lazo},\ and\ \citenamefont
  {Heinze}}]{Ferriani2010}%
  \BibitemOpen
  \bibfield  {author} {\bibinfo {author} {\bibfnamefont {P.}~\bibnamefont
  {Ferriani}}, \bibinfo {author} {\bibfnamefont {C.}~\bibnamefont {Lazo}},\
  and\ \bibinfo {author} {\bibfnamefont {S.}~\bibnamefont {Heinze}},\
  }\bibfield  {title} {\bibinfo {title} {Origin of the spin polarization of
  magnetic scanning tunneling microscopy tips},\ }\href
  {https://doi.org/10.1103/PhysRevB.82.054411} {\bibfield  {journal} {\bibinfo
  {journal} {Phys. Rev. B}\ }\textbf {\bibinfo {volume} {82}},\ \bibinfo
  {pages} {054411} (\bibinfo {year} {2010})}\BibitemShut {NoStop}%
\bibitem [{\citenamefont {Zhou}\ \emph {et~al.}(2010)\citenamefont {Zhou},
  \citenamefont {Meier}, \citenamefont {Wiebe},\ and\ \citenamefont
  {Wiesendanger}}]{Zhou2010}%
  \BibitemOpen
  \bibfield  {author} {\bibinfo {author} {\bibfnamefont {L.}~\bibnamefont
  {Zhou}}, \bibinfo {author} {\bibfnamefont {F.}~\bibnamefont {Meier}},
  \bibinfo {author} {\bibfnamefont {J.}~\bibnamefont {Wiebe}},\ and\ \bibinfo
  {author} {\bibfnamefont {R.}~\bibnamefont {Wiesendanger}},\ }\bibfield
  {title} {\bibinfo {title} {Inversion of spin polarization above individual
  magnetic adatoms},\ }\href {https://doi.org/10.1103/PhysRevB.82.012409}
  {\bibfield  {journal} {\bibinfo  {journal} {Phys. Rev. B}\ }\textbf {\bibinfo
  {volume} {82}},\ \bibinfo {pages} {012409} (\bibinfo {year}
  {2010})}\BibitemShut {NoStop}%
\bibitem [{\citenamefont {Fert}(2008)}]{Fert2008}%
  \BibitemOpen
  \bibfield  {author} {\bibinfo {author} {\bibfnamefont {A.}~\bibnamefont
  {Fert}},\ }\bibfield  {title} {\bibinfo {title} {The present and the future
  of spintronics},\ }\href
  {https://doi.org/https://doi.org/10.1016/j.tsf.2008.08.172} {\bibfield
  {journal} {\bibinfo  {journal} {Thin Solid Films}\ }\textbf {\bibinfo
  {volume} {517}},\ \bibinfo {pages} {2 } (\bibinfo {year} {2008})}\BibitemShut
  {NoStop}%
\bibitem [{\citenamefont {Sinova}\ and\ \citenamefont
  {{\v{Z}}uti{\'{c}}}(2012)}]{Sinova2012}%
  \BibitemOpen
  \bibfield  {author} {\bibinfo {author} {\bibfnamefont {J.}~\bibnamefont
  {Sinova}}\ and\ \bibinfo {author} {\bibfnamefont {I.}~\bibnamefont
  {{\v{Z}}uti{\'{c}}}},\ }\bibfield  {title} {\bibinfo {title} {{New moves of
  the spintronics tango}},\ }\href {https://doi.org/10.1038/nmat3304}
  {\bibfield  {journal} {\bibinfo  {journal} {Nat. Mater.}\ }\textbf {\bibinfo
  {volume} {11}},\ \bibinfo {pages} {368} (\bibinfo {year} {2012})}\BibitemShut
  {NoStop}%
\bibitem [{\citenamefont {Shick}\ \emph {et~al.}(2006)\citenamefont {Shick},
  \citenamefont {M\'aca}, \citenamefont {Ma\ifmmode~\check{s}\else
  \v{s}\fi{}ek},\ and\ \citenamefont {Jungwirth}}]{Shick2006}%
  \BibitemOpen
  \bibfield  {author} {\bibinfo {author} {\bibfnamefont {A.~B.}\ \bibnamefont
  {Shick}}, \bibinfo {author} {\bibfnamefont {F.}~\bibnamefont {M\'aca}},
  \bibinfo {author} {\bibfnamefont {J.}~\bibnamefont {Ma\ifmmode~\check{s}\else
  \v{s}\fi{}ek}},\ and\ \bibinfo {author} {\bibfnamefont {T.}~\bibnamefont
  {Jungwirth}},\ }\bibfield  {title} {\bibinfo {title} {Prospect for room
  temperature tunneling anisotropic magnetoresistance effect: Density of states
  anisotropies in $\mathrm{CoPt}$ systems},\ }\href
  {https://doi.org/10.1103/PhysRevB.73.024418} {\bibfield  {journal} {\bibinfo
  {journal} {Phys. Rev. B}\ }\textbf {\bibinfo {volume} {73}},\ \bibinfo
  {pages} {024418} (\bibinfo {year} {2006})}\BibitemShut {NoStop}%
\bibitem [{\citenamefont {Chantis}\ \emph {et~al.}(2007)\citenamefont
  {Chantis}, \citenamefont {Belashchenko}, \citenamefont {Tsymbal},\ and\
  \citenamefont {van Schilfgaarde}}]{Chantis2007}%
  \BibitemOpen
  \bibfield  {author} {\bibinfo {author} {\bibfnamefont {A.~N.}\ \bibnamefont
  {Chantis}}, \bibinfo {author} {\bibfnamefont {K.~D.}\ \bibnamefont
  {Belashchenko}}, \bibinfo {author} {\bibfnamefont {E.~Y.}\ \bibnamefont
  {Tsymbal}},\ and\ \bibinfo {author} {\bibfnamefont {M.}~\bibnamefont {van
  Schilfgaarde}},\ }\bibfield  {title} {\bibinfo {title} {Tunneling anisotropic
  magnetoresistance driven by resonant surface states: First-principles
  calculations on an {Fe(001)} surface},\ }\href
  {https://doi.org/10.1103/PhysRevLett.98.046601} {\bibfield  {journal}
  {\bibinfo  {journal} {Phys. Rev. Lett.}\ }\textbf {\bibinfo {volume} {98}},\
  \bibinfo {pages} {046601} (\bibinfo {year} {2007})}\BibitemShut {NoStop}%
\bibitem [{\citenamefont {Gould}\ \emph {et~al.}(2004)\citenamefont {Gould},
  \citenamefont {R\"uster}, \citenamefont {Jungwirth}, \citenamefont {Girgis},
  \citenamefont {Schott}, \citenamefont {Giraud}, \citenamefont {Brunner},
  \citenamefont {Schmidt},\ and\ \citenamefont {Molenkamp}}]{Gould2004}%
  \BibitemOpen
  \bibfield  {author} {\bibinfo {author} {\bibfnamefont {C.}~\bibnamefont
  {Gould}}, \bibinfo {author} {\bibfnamefont {C.}~\bibnamefont {R\"uster}},
  \bibinfo {author} {\bibfnamefont {T.}~\bibnamefont {Jungwirth}}, \bibinfo
  {author} {\bibfnamefont {E.}~\bibnamefont {Girgis}}, \bibinfo {author}
  {\bibfnamefont {G.~M.}\ \bibnamefont {Schott}}, \bibinfo {author}
  {\bibfnamefont {R.}~\bibnamefont {Giraud}}, \bibinfo {author} {\bibfnamefont
  {K.}~\bibnamefont {Brunner}}, \bibinfo {author} {\bibfnamefont
  {G.}~\bibnamefont {Schmidt}},\ and\ \bibinfo {author} {\bibfnamefont {L.~W.}\
  \bibnamefont {Molenkamp}},\ }\bibfield  {title} {\bibinfo {title} {Tunneling
  anisotropic magnetoresistance: A spin-valve-like tunnel magnetoresistance
  using a single magnetic layer},\ }\href
  {https://doi.org/10.1103/PhysRevLett.93.117203} {\bibfield  {journal}
  {\bibinfo  {journal} {Phys. Rev. Lett.}\ }\textbf {\bibinfo {volume} {93}},\
  \bibinfo {pages} {117203} (\bibinfo {year} {2004})}\BibitemShut {NoStop}%
\bibitem [{\citenamefont {Matos-Abiague}\ and\ \citenamefont
  {Fabian}(2009)}]{Matos2009a}%
  \BibitemOpen
  \bibfield  {author} {\bibinfo {author} {\bibfnamefont {A.}~\bibnamefont
  {Matos-Abiague}}\ and\ \bibinfo {author} {\bibfnamefont {J.}~\bibnamefont
  {Fabian}},\ }\bibfield  {title} {\bibinfo {title} {Anisotropic tunneling
  magnetoresistance and tunneling anisotropic magnetoresistance: Spin-orbit
  coupling in magnetic tunnel junctions},\ }\href
  {https://doi.org/10.1103/PhysRevB.79.155303} {\bibfield  {journal} {\bibinfo
  {journal} {Phys. Rev. B}\ }\textbf {\bibinfo {volume} {79}},\ \bibinfo
  {pages} {155303} (\bibinfo {year} {2009})}\BibitemShut {NoStop}%
\bibitem [{\citenamefont {Matos-Abiague}\ \emph {et~al.}(2009)\citenamefont
  {Matos-Abiague}, \citenamefont {Gmitra},\ and\ \citenamefont
  {Fabian}}]{Matos2009b}%
  \BibitemOpen
  \bibfield  {author} {\bibinfo {author} {\bibfnamefont {A.}~\bibnamefont
  {Matos-Abiague}}, \bibinfo {author} {\bibfnamefont {M.}~\bibnamefont
  {Gmitra}},\ and\ \bibinfo {author} {\bibfnamefont {J.}~\bibnamefont
  {Fabian}},\ }\bibfield  {title} {\bibinfo {title} {Angular dependence of the
  tunneling anisotropic magnetoresistance in magnetic tunnel junctions},\
  }\href {https://doi.org/10.1103/PhysRevB.80.045312} {\bibfield  {journal}
  {\bibinfo  {journal} {Phys. Rev. B}\ }\textbf {\bibinfo {volume} {80}},\
  \bibinfo {pages} {045312} (\bibinfo {year} {2009})}\BibitemShut {NoStop}%
\bibitem [{\citenamefont {Gao}\ \emph {et~al.}(2007)\citenamefont {Gao},
  \citenamefont {Jiang}, \citenamefont {Yang}, \citenamefont {Burton},
  \citenamefont {Tsymbal},\ and\ \citenamefont {Parkin}}]{Gao2007}%
  \BibitemOpen
  \bibfield  {author} {\bibinfo {author} {\bibfnamefont {L.}~\bibnamefont
  {Gao}}, \bibinfo {author} {\bibfnamefont {X.}~\bibnamefont {Jiang}}, \bibinfo
  {author} {\bibfnamefont {S.-H.}\ \bibnamefont {Yang}}, \bibinfo {author}
  {\bibfnamefont {J.~D.}\ \bibnamefont {Burton}}, \bibinfo {author}
  {\bibfnamefont {E.~Y.}\ \bibnamefont {Tsymbal}},\ and\ \bibinfo {author}
  {\bibfnamefont {S.~S.~P.}\ \bibnamefont {Parkin}},\ }\bibfield  {title}
  {\bibinfo {title} {Bias voltage dependence of tunneling anisotropic
  magnetoresistance in magnetic tunnel junctions with {MgO} and
  {${\mathrm{Al}}_{2}{\mathrm{O}}_{3}$} tunnel barriers},\ }\href
  {https://doi.org/10.1103/PhysRevLett.99.226602} {\bibfield  {journal}
  {\bibinfo  {journal} {Phys. Rev. Lett.}\ }\textbf {\bibinfo {volume} {99}},\
  \bibinfo {pages} {226602} (\bibinfo {year} {2007})}\BibitemShut {NoStop}%
\bibitem [{\citenamefont {Viret}\ \emph {et~al.}(2006)\citenamefont {Viret},
  \citenamefont {Gabureac}, \citenamefont {Ott}, \citenamefont {Fermon},
  \citenamefont {Barreteau}, \citenamefont {Autes},\ and\ \citenamefont
  {Guirado-Lopez}}]{Viret2006}%
  \BibitemOpen
  \bibfield  {author} {\bibinfo {author} {\bibfnamefont {M.}~\bibnamefont
  {Viret}}, \bibinfo {author} {\bibfnamefont {M.}~\bibnamefont {Gabureac}},
  \bibinfo {author} {\bibfnamefont {F.}~\bibnamefont {Ott}}, \bibinfo {author}
  {\bibfnamefont {C.}~\bibnamefont {Fermon}}, \bibinfo {author} {\bibfnamefont
  {C.}~\bibnamefont {Barreteau}}, \bibinfo {author} {\bibfnamefont
  {G.}~\bibnamefont {Autes}},\ and\ \bibinfo {author} {\bibfnamefont
  {R.}~\bibnamefont {Guirado-Lopez}},\ }\bibfield  {title} {\bibinfo {title}
  {Giant anisotropic magneto-resistance in ferromagnetic atomic contacts},\
  }\href {https://doi.org/10.1140/epjb/e2006-00201-3} {\bibfield  {journal}
  {\bibinfo  {journal} {The European Physical Journal B - Condensed Matter and
  Complex Systems}\ }\textbf {\bibinfo {volume} {51}},\ \bibinfo {pages} {1}
  (\bibinfo {year} {2006})}\BibitemShut {NoStop}%
\bibitem [{\citenamefont {Bolotin}\ \emph {et~al.}(2006)\citenamefont
  {Bolotin}, \citenamefont {Kuemmeth},\ and\ \citenamefont
  {Ralph}}]{Bolotin2006}%
  \BibitemOpen
  \bibfield  {author} {\bibinfo {author} {\bibfnamefont {K.~I.}\ \bibnamefont
  {Bolotin}}, \bibinfo {author} {\bibfnamefont {F.}~\bibnamefont {Kuemmeth}},\
  and\ \bibinfo {author} {\bibfnamefont {D.~C.}\ \bibnamefont {Ralph}},\
  }\bibfield  {title} {\bibinfo {title} {Anisotropic magnetoresistance and
  anisotropic tunneling magnetoresistance due to quantum interference in
  ferromagnetic metal break junctions},\ }\href
  {https://doi.org/10.1103/PhysRevLett.97.127202} {\bibfield  {journal}
  {\bibinfo  {journal} {Phys. Rev. Lett.}\ }\textbf {\bibinfo {volume} {97}},\
  \bibinfo {pages} {127202} (\bibinfo {year} {2006})}\BibitemShut {NoStop}%
\bibitem [{\citenamefont {N\'eel}\ \emph {et~al.}(2013)\citenamefont {N\'eel},
  \citenamefont {Schr\"oder}, \citenamefont {Ruppelt}, \citenamefont
  {Ferriani}, \citenamefont {Kr\"oger}, \citenamefont {Berndt},\ and\
  \citenamefont {Heinze}}]{Neel2013}%
  \BibitemOpen
  \bibfield  {author} {\bibinfo {author} {\bibfnamefont {N.}~\bibnamefont
  {N\'eel}}, \bibinfo {author} {\bibfnamefont {S.}~\bibnamefont {Schr\"oder}},
  \bibinfo {author} {\bibfnamefont {N.}~\bibnamefont {Ruppelt}}, \bibinfo
  {author} {\bibfnamefont {P.}~\bibnamefont {Ferriani}}, \bibinfo {author}
  {\bibfnamefont {J.}~\bibnamefont {Kr\"oger}}, \bibinfo {author}
  {\bibfnamefont {R.}~\bibnamefont {Berndt}},\ and\ \bibinfo {author}
  {\bibfnamefont {S.}~\bibnamefont {Heinze}},\ }\bibfield  {title} {\bibinfo
  {title} {Tunneling anisotropic magnetoresistance at the single-atom limit},\
  }\href {https://doi.org/10.1103/PhysRevLett.110.037202} {\bibfield  {journal}
  {\bibinfo  {journal} {Phys. Rev. Lett.}\ }\textbf {\bibinfo {volume} {110}},\
  \bibinfo {pages} {037202} (\bibinfo {year} {2013})}\BibitemShut {NoStop}%
\bibitem [{\citenamefont {Herv\'e}\ \emph {et~al.}(2018)\citenamefont
  {Herv\'e}, \citenamefont {Balashov}, \citenamefont {Ernst},\ and\
  \citenamefont {Wulfhekel}}]{Herve2018b}%
  \BibitemOpen
  \bibfield  {author} {\bibinfo {author} {\bibfnamefont {M.}~\bibnamefont
  {Herv\'e}}, \bibinfo {author} {\bibfnamefont {T.}~\bibnamefont {Balashov}},
  \bibinfo {author} {\bibfnamefont {A.}~\bibnamefont {Ernst}},\ and\ \bibinfo
  {author} {\bibfnamefont {W.}~\bibnamefont {Wulfhekel}},\ }\bibfield  {title}
  {\bibinfo {title} {Large tunneling anisotropic magnetoresistance mediated by
  surface states},\ }\href {https://doi.org/10.1103/PhysRevB.97.220406}
  {\bibfield  {journal} {\bibinfo  {journal} {Phys. Rev. B}\ }\textbf {\bibinfo
  {volume} {97}},\ \bibinfo {pages} {220406(R)} (\bibinfo {year}
  {2018})}\BibitemShut {NoStop}%
\bibitem [{\citenamefont {Sch{\"{o}}neberg}\ \emph {et~al.}(2018)\citenamefont
  {Sch{\"{o}}neberg}, \citenamefont {Ferriani}, \citenamefont {Heinze},
  \citenamefont {Weismann},\ and\ \citenamefont {Berndt}}]{Schoneberg2018}%
  \BibitemOpen
  \bibfield  {author} {\bibinfo {author} {\bibfnamefont {J.}~\bibnamefont
  {Sch{\"{o}}neberg}}, \bibinfo {author} {\bibfnamefont {P.}~\bibnamefont
  {Ferriani}}, \bibinfo {author} {\bibfnamefont {S.}~\bibnamefont {Heinze}},
  \bibinfo {author} {\bibfnamefont {A.}~\bibnamefont {Weismann}},\ and\
  \bibinfo {author} {\bibfnamefont {R.}~\bibnamefont {Berndt}},\ }\bibfield
  {title} {\bibinfo {title} {{Tunneling anisotropic magnetoresistance via
  molecular $\pi$ orbitals of Pb dimers}},\ }\href
  {https://doi.org/10.1103/PhysRevB.97.041114} {\bibfield  {journal} {\bibinfo
  {journal} {Phys. Rev. B}\ }\textbf {\bibinfo {volume} {97}},\ \bibinfo
  {pages} {041114(R)} (\bibinfo {year} {2018})}\BibitemShut {NoStop}%
\bibitem [{\citenamefont {Haldar}\ \emph {et~al.}(2019)\citenamefont {Haldar},
  \citenamefont {Gutzeit},\ and\ \citenamefont {Heinze}}]{Haldar2019}%
  \BibitemOpen
  \bibfield  {author} {\bibinfo {author} {\bibfnamefont {S.}~\bibnamefont
  {Haldar}}, \bibinfo {author} {\bibfnamefont {M.}~\bibnamefont {Gutzeit}},\
  and\ \bibinfo {author} {\bibfnamefont {S.}~\bibnamefont {Heinze}},\
  }\bibfield  {title} {\bibinfo {title} {Tunneling anisotropic
  magnetoresistance of {Pb} and {Bi} adatoms and dimers on {Mn/W(110)}: A
  first-principles study},\ }\href
  {https://doi.org/10.1103/PhysRevB.100.094412} {\bibfield  {journal} {\bibinfo
   {journal} {Phys. Rev. B}\ }\textbf {\bibinfo {volume} {100}},\ \bibinfo
  {pages} {094412} (\bibinfo {year} {2019})}\BibitemShut {NoStop}%
\bibitem [{\citenamefont {Caffrey}\ \emph {et~al.}(2014)\citenamefont
  {Caffrey}, \citenamefont {Schr{\"{o}}der}, \citenamefont {Ferriani},\ and\
  \citenamefont {Heinze}}]{Caffrey2014a}%
  \BibitemOpen
  \bibfield  {author} {\bibinfo {author} {\bibfnamefont {N.~M.}\ \bibnamefont
  {Caffrey}}, \bibinfo {author} {\bibfnamefont {S.}~\bibnamefont
  {Schr{\"{o}}der}}, \bibinfo {author} {\bibfnamefont {P.}~\bibnamefont
  {Ferriani}},\ and\ \bibinfo {author} {\bibfnamefont {S.}~\bibnamefont
  {Heinze}},\ }\bibfield  {title} {\bibinfo {title} {{Tunneling anisotropic
  magnetoresistance effect of single adatoms on a noncollinear magnetic
  surface}},\ }\href {https://doi.org/10.1088/0953-8984/26/39/394010}
  {\bibfield  {journal} {\bibinfo  {journal} {Journal of Physics: Condensed
  Matter}\ }\textbf {\bibinfo {volume} {26}},\ \bibinfo {pages} {394010}
  (\bibinfo {year} {2014})}\BibitemShut {NoStop}%
\bibitem [{\citenamefont {Heinze}\ \emph {et~al.}(2000)\citenamefont {Heinze},
  \citenamefont {Bode}, \citenamefont {Kubetzka}, \citenamefont {Pietzsch},
  \citenamefont {Nie}, \citenamefont {Bl{\"u}gel},\ and\ \citenamefont
  {Wiesendanger}}]{Heinze2000}%
  \BibitemOpen
  \bibfield  {author} {\bibinfo {author} {\bibfnamefont {S.}~\bibnamefont
  {Heinze}}, \bibinfo {author} {\bibfnamefont {M.}~\bibnamefont {Bode}},
  \bibinfo {author} {\bibfnamefont {A.}~\bibnamefont {Kubetzka}}, \bibinfo
  {author} {\bibfnamefont {O.}~\bibnamefont {Pietzsch}}, \bibinfo {author}
  {\bibfnamefont {X.}~\bibnamefont {Nie}}, \bibinfo {author} {\bibfnamefont
  {S.}~\bibnamefont {Bl{\"u}gel}},\ and\ \bibinfo {author} {\bibfnamefont
  {R.}~\bibnamefont {Wiesendanger}},\ }\bibfield  {title} {\bibinfo {title}
  {Real-space imaging of two-dimensional antiferromagnetism on the atomic
  scale},\ }\href {https://doi.org/10.1126/science.288.5472.1805} {\bibfield
  {journal} {\bibinfo  {journal} {Science}\ }\textbf {\bibinfo {volume}
  {288}},\ \bibinfo {pages} {1805} (\bibinfo {year} {2000})}\BibitemShut
  {NoStop}%
\bibitem [{\citenamefont {Bl\"ochl}(1994)}]{blo}%
  \BibitemOpen
  \bibfield  {author} {\bibinfo {author} {\bibfnamefont {P.~E.}\ \bibnamefont
  {Bl\"ochl}},\ }\bibfield  {title} {\bibinfo {title} {Projector augmented-wave
  method},\ }\href {https://doi.org/10.1103/PhysRevB.50.17953} {\bibfield
  {journal} {\bibinfo  {journal} {Phys. Rev. B}\ }\textbf {\bibinfo {volume}
  {50}},\ \bibinfo {pages} {17953} (\bibinfo {year} {1994})}\BibitemShut
  {NoStop}%
\bibitem [{\citenamefont {Kresse}\ and\ \citenamefont {Joubert}(1999)}]{blo1}%
  \BibitemOpen
  \bibfield  {author} {\bibinfo {author} {\bibfnamefont {G.}~\bibnamefont
  {Kresse}}\ and\ \bibinfo {author} {\bibfnamefont {D.}~\bibnamefont
  {Joubert}},\ }\bibfield  {title} {\bibinfo {title} {From ultrasoft
  pseudopotentials to the projector augmented-wave method},\ }\href
  {https://doi.org/10.1103/PhysRevB.59.1758} {\bibfield  {journal} {\bibinfo
  {journal} {Phys. Rev. B}\ }\textbf {\bibinfo {volume} {59}},\ \bibinfo
  {pages} {1758} (\bibinfo {year} {1999})}\BibitemShut {NoStop}%
\bibitem [{\citenamefont {Kresse}\ and\ \citenamefont
  {Furthm\"uller}(1996)}]{vasp1}%
  \BibitemOpen
  \bibfield  {author} {\bibinfo {author} {\bibfnamefont {G.}~\bibnamefont
  {Kresse}}\ and\ \bibinfo {author} {\bibfnamefont {J.}~\bibnamefont
  {Furthm\"uller}},\ }\bibfield  {title} {\bibinfo {title} {Efficient iterative
  schemes for \textit{ab Initio} total-energy calculations using a plane-wave
  basis set},\ }\href {https://doi.org/10.1103/PhysRevB.54.11169} {\bibfield
  {journal} {\bibinfo  {journal} {Phys. Rev. B}\ }\textbf {\bibinfo {volume}
  {54}},\ \bibinfo {pages} {11169} (\bibinfo {year} {1996})}\BibitemShut
  {NoStop}%
\bibitem [{vas()}]{vasp2}%
  \BibitemOpen
  \href@noop {} {}\bibinfo {howpublished} {See
  \url{https://www.vasp.at}}\BibitemShut {NoStop}%
\bibitem [{\citenamefont {Perdew}\ \emph {et~al.}(1996)\citenamefont {Perdew},
  \citenamefont {Burke},\ and\ \citenamefont {Ernzerhof}}]{PBE}%
  \BibitemOpen
  \bibfield  {author} {\bibinfo {author} {\bibfnamefont {J.~P.}\ \bibnamefont
  {Perdew}}, \bibinfo {author} {\bibfnamefont {K.}~\bibnamefont {Burke}},\ and\
  \bibinfo {author} {\bibfnamefont {M.}~\bibnamefont {Ernzerhof}},\ }\bibfield
  {title} {\bibinfo {title} {Generalized gradient approximation made simple},\
  }\href {https://doi.org/10.1103/PhysRevLett.77.3865} {\bibfield  {journal}
  {\bibinfo  {journal} {Phys. Rev. Lett.}\ }\textbf {\bibinfo {volume} {77}},\
  \bibinfo {pages} {3865} (\bibinfo {year} {1996})}\BibitemShut {NoStop}%
\bibitem [{\citenamefont {Perdew}\ \emph {et~al.}(1997)\citenamefont {Perdew},
  \citenamefont {Burke},\ and\ \citenamefont {Ernzerhof}}]{PBEerr}%
  \BibitemOpen
  \bibfield  {author} {\bibinfo {author} {\bibfnamefont {J.~P.}\ \bibnamefont
  {Perdew}}, \bibinfo {author} {\bibfnamefont {K.}~\bibnamefont {Burke}},\ and\
  \bibinfo {author} {\bibfnamefont {M.}~\bibnamefont {Ernzerhof}},\ }\bibfield
  {title} {\bibinfo {title} {Generalized gradient approximation made simple
  [{Phys. Rev. Lett. \textbf{77}, 3865 (1996)}]},\ }\href
  {https://doi.org/10.1103/PhysRevLett.78.1396} {\bibfield  {journal} {\bibinfo
   {journal} {Phys. Rev. Lett.}\ }\textbf {\bibinfo {volume} {78}},\ \bibinfo
  {pages} {1396} (\bibinfo {year} {1997})}\BibitemShut {NoStop}%
\bibitem [{\citenamefont {Monkhorst}\ and\ \citenamefont
  {Pack}(1976)}]{Monkhorst1976}%
  \BibitemOpen
  \bibfield  {author} {\bibinfo {author} {\bibfnamefont {H.~J.}\ \bibnamefont
  {Monkhorst}}\ and\ \bibinfo {author} {\bibfnamefont {J.~D.}\ \bibnamefont
  {Pack}},\ }\bibfield  {title} {\bibinfo {title} {{Special points for
  Brillouin-zone integrations}},\ }\href
  {https://doi.org/10.1103/PhysRevB.13.5188} {\bibfield  {journal} {\bibinfo
  {journal} {Phys. Rev. B}\ }\textbf {\bibinfo {volume} {13}},\ \bibinfo
  {pages} {5188} (\bibinfo {year} {1976})}\BibitemShut {NoStop}%
\bibitem [{\citenamefont {Hobbs}\ \emph {et~al.}(2000)\citenamefont {Hobbs},
  \citenamefont {Kresse},\ and\ \citenamefont {Hafner}}]{Hobbs2000}%
  \BibitemOpen
  \bibfield  {author} {\bibinfo {author} {\bibfnamefont {D.}~\bibnamefont
  {Hobbs}}, \bibinfo {author} {\bibfnamefont {G.}~\bibnamefont {Kresse}},\ and\
  \bibinfo {author} {\bibfnamefont {J.}~\bibnamefont {Hafner}},\ }\bibfield
  {title} {\bibinfo {title} {{Fully unconstrained noncollinear magnetism within
  the projector augmented-wave method}},\ }\href
  {https://doi.org/10.1103/PhysRevB.62.11556} {\bibfield  {journal} {\bibinfo
  {journal} {Phys. Rev. B}\ }\textbf {\bibinfo {volume} {62}},\ \bibinfo
  {pages} {11556} (\bibinfo {year} {2000})}\BibitemShut {NoStop}%
\bibitem [{\citenamefont {Bode}\ \emph {et~al.}(1999)\citenamefont {Bode},
  \citenamefont {Hennefarth}, \citenamefont {Haude}, \citenamefont {Getzlaff},\
  and\ \citenamefont {Wiesendanger}}]{Bode1998}%
  \BibitemOpen
  \bibfield  {author} {\bibinfo {author} {\bibfnamefont {M.}~\bibnamefont
  {Bode}}, \bibinfo {author} {\bibfnamefont {M.}~\bibnamefont {Hennefarth}},
  \bibinfo {author} {\bibfnamefont {D.}~\bibnamefont {Haude}}, \bibinfo
  {author} {\bibfnamefont {M.}~\bibnamefont {Getzlaff}},\ and\ \bibinfo
  {author} {\bibfnamefont {R.}~\bibnamefont {Wiesendanger}},\ }\bibfield
  {title} {\bibinfo {title} {Growth of thin {Mn} films on {W(110)} studied by
  means of in-situ scanning tunnelling microscopy},\ }\href
  {https://doi.org/https://doi.org/10.1016/S0039-6028(99)00447-1} {\bibfield
  {journal} {\bibinfo  {journal} {Surface Science}\ }\textbf {\bibinfo {volume}
  {432}},\ \bibinfo {pages} {8 } (\bibinfo {year} {1999})}\BibitemShut
  {NoStop}%
\bibitem [{Note1()}]{Note1}%
  \BibitemOpen
  \bibinfo {note} {This quantity can also be evaluated directly at the
  adatom}\BibitemShut {NoStop}%
\bibitem [{\citenamefont {Tersoff}\ and\ \citenamefont
  {Hamann}(1983)}]{Tersoff1983}%
  \BibitemOpen
  \bibfield  {author} {\bibinfo {author} {\bibfnamefont {J.}~\bibnamefont
  {Tersoff}}\ and\ \bibinfo {author} {\bibfnamefont {D.~R.}\ \bibnamefont
  {Hamann}},\ }\bibfield  {title} {\bibinfo {title} {Theory and application for
  the scanning tunneling microscope},\ }\href
  {https://doi.org/10.1103/PhysRevLett.50.1998} {\bibfield  {journal} {\bibinfo
   {journal} {Phys. Rev. Lett.}\ }\textbf {\bibinfo {volume} {50}},\ \bibinfo
  {pages} {1998} (\bibinfo {year} {1983})}\BibitemShut {NoStop}%
\bibitem [{\citenamefont {Tersoff}\ and\ \citenamefont
  {Hamann}(1985)}]{Tersoff1985}%
  \BibitemOpen
  \bibfield  {author} {\bibinfo {author} {\bibfnamefont {J.}~\bibnamefont
  {Tersoff}}\ and\ \bibinfo {author} {\bibfnamefont {D.~R.}\ \bibnamefont
  {Hamann}},\ }\bibfield  {title} {\bibinfo {title} {Theory of the scanning
  tunneling microscope},\ }\href {https://doi.org/10.1103/PhysRevB.31.805}
  {\bibfield  {journal} {\bibinfo  {journal} {Phys. Rev. B}\ }\textbf {\bibinfo
  {volume} {31}},\ \bibinfo {pages} {805} (\bibinfo {year} {1985})}\BibitemShut
  {NoStop}%
\bibitem [{\citenamefont {Wortmann}\ \emph {et~al.}(2001)\citenamefont
  {Wortmann}, \citenamefont {Heinze}, \citenamefont {Kurz}, \citenamefont
  {Bihlmayer},\ and\ \citenamefont {Bl{\"{u}}gel}}]{Wortmann2001}%
  \BibitemOpen
  \bibfield  {author} {\bibinfo {author} {\bibfnamefont {D.}~\bibnamefont
  {Wortmann}}, \bibinfo {author} {\bibfnamefont {S.}~\bibnamefont {Heinze}},
  \bibinfo {author} {\bibfnamefont {P.}~\bibnamefont {Kurz}}, \bibinfo {author}
  {\bibfnamefont {G.}~\bibnamefont {Bihlmayer}},\ and\ \bibinfo {author}
  {\bibfnamefont {S.}~\bibnamefont {Bl{\"{u}}gel}},\ }\bibfield  {title}
  {\bibinfo {title} {Resolving complex atomic-scale spin structures by
  spin-polarized scanning tunneling microscopy},\ }\href
  {https://doi.org/10.1103/PhysRevLett.86.4132} {\bibfield  {journal} {\bibinfo
   {journal} {Phys. Rev. Lett.}\ }\textbf {\bibinfo {volume} {86}},\ \bibinfo
  {pages} {4132} (\bibinfo {year} {2001})}\BibitemShut {NoStop}%
\bibitem [{Note2()}]{Note2}%
  \BibitemOpen
  \bibinfo {note} {\protect \leavevmode {\protect It is to be
  noted that in our calculations, the magnetization directions are applied to
  the complete system consisting of the adsorbed adatom and the underlying
  Mn/W(110) surface.}}\BibitemShut {Stop}%
\bibitem [{\citenamefont {Abate}\ and\ \citenamefont
  {Asdente}(1969)}]{Abate1965}%
  \BibitemOpen
  \bibfield  {author} {\bibinfo {author} {\bibfnamefont {E.}~\bibnamefont
  {Abate}}\ and\ \bibinfo {author} {\bibfnamefont {M.}~\bibnamefont
  {Asdente}},\ }\bibfield  {title} {\bibinfo {title} {Tight-binding calculation
  of $3d$ bands of fe with and without spin-orbit coupling},\ }\href@noop {}
  {\bibfield  {journal} {\bibinfo  {journal} {Phys. Rev.}\ }\textbf {\bibinfo
  {volume} {185}},\ \bibinfo {pages} {861} (\bibinfo {year}
  {1969})}\BibitemShut {NoStop}%
\bibitem [{\citenamefont {St{\"o}hr}\ and\ \citenamefont
  {Siegmann}(2006)}]{StoehrSiegmann}%
  \BibitemOpen
  \bibfield  {author} {\bibinfo {author} {\bibfnamefont {J.}~\bibnamefont
  {St{\"o}hr}}\ and\ \bibinfo {author} {\bibfnamefont {H.~C.}\ \bibnamefont
  {Siegmann}},\ }\href@noop {} {\emph {\bibinfo {title} {Magnetism - From
  Fundamentals to Nanoscale Dynamics}}}\ (\bibinfo  {publisher}
  {Springer-Verlag Berlin Heidelberg},\ \bibinfo {year} {2006})\BibitemShut
  {NoStop}%
\end{thebibliography}
